\documentclass[aps,prd,preprintnumbers,amsmath,amssymb,nofootinbib,10pt]{revtex4}

\pdfoutput=1
\usepackage{bm}
\usepackage{graphicx}
\usepackage{subfig}
\usepackage{array}
\usepackage{color}
\usepackage{ulem}
\usepackage{amssymb}
\usepackage{graphicx}
\usepackage{caption}
\usepackage{multirow}
\usepackage{slashed}
\usepackage{hyperref}

\begin{document}

\title{Tau reconstruction methods at an electron-positron collider in the search for new physics}

\author{Jinmian Li}\email{jinmian.li@adelaide.edu.au}\quad
\author{Anthony G.\ Williams}\email{anthony.williams@adelaide.edu.au}

\affiliation{ARC Centre of Excellence for Particle Physics at the Terascale and CSSM, Department of Physics,
University of Adelaide, Adelaide, SA 5005, Australia}

\begin{abstract}
By exploiting the relatively long lifetime of the tau lepton, we propose several novel methods for searching for new physics at an electron-positron collider. We consider processes that involve final states consisting of a tau lepton pair plus two missing particles. The mass and spin of the new physics particles can be measured in 3-prong tau decays. The tau polarization, which reflects the coupling to new physics, can be measured from the $\tau \to \pi \nu$ decay channel using the impact parameter distribution of the charged pion. We also discuss the corresponding backgrounds for these measurements, the next-to-leading order (NLO) effects, and the implications of finite detector resolution.  
\end{abstract}

\maketitle

\section{Introduction}
Despite the excellent successes of the Standard Model (SM) of particle physics it is clearly incomplete, e.g., it does not explain Dark Matter (DM),
the gauge hierarchy problem or the current excess of matter over antimatter in the universe.
Supersymmetry (SUSY)~\cite{Nilles19841,Haber198575} is one of the leading candidate theories for new physics, i.e.,
physics theories that go Beyond the Standard Model (BSM). If supersymmetric particles (sparticles) have masses below the TeV scale then the gauge hierarchy problem in the SM is solved naturally. In supersymmetric extensions of the SM (SSM), the gauge couplings are unified at a high energy scale around $\mathcal{O}(10^{16})$ GeV~\cite{Ellis:1990zq,Amaldi:1991cn,Langacker:1991an}, which is a strong indication of the possible existence of an overarching Grand Unified Theory (GUT). With the assumed existence of a conserved R-parity in these SSMs the lightest supersymmetric particle (LSP) become stable and so serves as a potential DM candidate~\cite{Goldberg:1983nd}. 

Considerable effort has gone into searches for evidence of SUSY at the Large Hadron Collider (LHC) at CERN. The lack of any evidence for
SUSY to date has pushed the potential
masses of sparticles to higher scales, which gives rise to the so-called little hierarchy problem~\cite{BasteroGil:2000bw,Bazzocchi:2012de}
for SUSY. At the 8 TeV LHC with an integrated luminosity of $\sim 20$ fb$^{-1}$ a gluino mass below 1.3 TeV and the first two generation squark masses below
850 GeV have been excluded with a 95\% Confidence Level (C.L.) in simplified models~\cite{Aad:2014wea,Khachatryan:2015vra}.
Third generation squarks have looser constraints, primarily because the proton contains no heavy valence quarks and
because the SM $t\bar{t}$ background is large. The
exclusion bounds have gone up to $\lesssim$ 700 GeV for both stop~\cite{Aad:2014bva,CMS-PAS-SUS-14-011} and sbottom~\cite{Aad:2013ija,CMS-PAS-SUS-13-018} when the LSP is light. 
Sparticles in the electro-weak sector can still be relatively light~\cite{Cheng:2012np}, since the current searches at the LHC can only exclude left-handed selectron and  smuon masses that are $\lesssim 300$ GeV~\cite{Khachatryan:2014qwa,Aad:2014vma}. There is no current LHC bound for the tau slepton (the stau)~\cite{Aad:2015eda}. The strongest bound for the stau is from Large Electron-Positron Collider (LEP) searches is relatively modest and is $m_{\tilde{\tau}} \gtrsim 90$ GeV~\cite{Heister:2001nk}.   
The possibility of a light stau is well motivated in many SUSY model frameworks. 
In Generalized Minimal Supergravity models~\cite{Li:2010xr,Balazs:2010ha} sleptons are much lighter than squarks due to the lack of an $SU(3)_c$ coupling. Furthermore, the large Yukawa coupling of the tau lepton makes the lighter stau even lighter
as it causes the stau mass to decrease more rapidly than the first two generation sleptons
when evolving down from the GUT scale.

The main reasons for the modest bounds on the stau are its small production cross section and the relatively large backgrounds at proton-proton colliders. A  future $e^+ e^-$ collider, where the processes are dominated by electroweak coupling, will provide a much improved environment for stau searches and would provide an ideal tool for the study of the electro-weak sector of SSMs.  Indeed, one of the main tasks for the proposed International Linear Collider (ILC)~\cite{Abe:2010aa,Baer:2013cma,Behnke:2013lya} is to provide excellent track reconstruction with fine momentum and impact parameter resolution in order to have a better measurement of Higgs couplings, such as $h \tau \tau$. The impact parameter resolution~\cite{Gaede:2014aza} is expected to be
$\sigma_{r \phi} = 5 ~\mu \text{m} \oplus [10 /(p/\text{GeV}) \sin^{3/2} \theta]~\mu \text{m}$~\footnote{We use the standard notation that 
$x\oplus y\equiv\sqrt{x^2+y^2}$ and $(p/\text{GeV})$ means, e.g., $6.5$ if the momentum is $p=6.5GeV$},
where $\theta$ is the angle between the lepton and the beam axis in the lab frame.
Note that such precision can only be reached in the $r\phi$ plane, since the nominal size~\cite{2009NIMPA.608..367I,Phinney:2007gp} of the beam bunch at 1 TeV  is $639 ~\text{nm (width)} \times 5.7 ~\text{nm (height)} \times 300 ~\mu \text{m (length)}$, where the length in the $z$-direction is much larger than the width and height.  

Following the next LHC run it may well be that the stau (and any new particles with similar decay channels) remains
unobserved even if it lies within the reach of a future $e^+ e^-$ collider. 
Possible new physics processes with electron and muon final states have been studied extensively in the context of an ILC and for the LHC~\cite{Gedalia:2009ym,Horton:2010bg,MoortgatPick:2011ix,Asano:2011aj,Moortgat-Picka:2015yla}. The energy distribution of the final state leptons can be used to measure the sleptons' mass precisely~\cite{Tsukamoto:1993gt}, while the threshold behavior of the excitation curves and the angular distribution of the observable particles can be used to determinate the spin of new physics particles~\cite{Choi:2006mr}. 
Final states involving the tau, while more complicated to analyze, are interesting and potentially very informative.
The end points of the tau jet energy distribution can be used directly to extract the masses in the decay chain~\cite{Nojiri:1996fp}, while the $E_{\pi}$ distribution of the $\tau \to \pi \nu$ channel and the $E_{\pi^{\pm}}/E_{\rho^{\pm}}$ distribution of the
$\tau^{\pm} \to \rho^{\pm} \nu$ channel are able to measure the tau polarization~\cite{Hagiwara:1989fn,Nojiri:274380,Boos:2003vf,Bechtle:2009em,Schade:2009zz,Berggren:2015qua}. Furthermore, the spin correlation of two tau leptons in the
$H \to \tau \tau$ channel can be used to determine the CP properties of the Higgs
boson~\cite{Desch:2003mw,Rouge:2005iy,Berge:2008dr,Berge:2013jra}. 
Some recent studies also show that vertex information for the tau decay can help in the reconstruction of the $H \to \tau \tau$ ~\cite{Desch:2003mw,Gripaios:2012th} process and other SM processes~\cite{Jeans:2015vaa} . 

In this study, we will take advantage of the anticipated powerful track reconstruction capabilities of the ILC to extract 
additional information about the final states of the tau decay and use this to assist in event reconstruction and the search for new physics, e.g., stau pair production with a subsequent decays of the form $\tilde{\tau} \to \tau \tilde{\chi}^0$. 
The discovery prospect of a 1 TeV ILC is analyzed here for a relatively heavy benchmark point of $m_{\tilde{\tau}} =300$ GeV and $m_{\tilde{\chi}^0}= 50$ GeV. However, our method will also be useful at other collision energies.
Reducing the collision energy while staying in the kinematically accessible region would increase the production rate of our signal. 

The paper is organised as follows. 
In Sec.~\ref{sec:tau} we give a brief overview of tau decay.
In Sec.~\ref{sec:npr} we present the theoretical framework for the mass and spin reconstruction and we
propose and discuss some techniques for the determination of tau polarization. The parton level results as well as
their statistical uncertainties are also
discussed. A more detailed analysis including detector effects is presented in Sec.~\ref{sec:num}
and finally we discuss our results and give our conclusions in Sec.~\ref{sec:conc}

\section{Tau decays}
\label{sec:tau}
The tau lepton has complicated decay branching fractions. Its dominant decay channels and their branching ratios~\cite{Agashe:2014kda} are given in Table~\ref{taubr}. The $2\pi$ and $3\pi$ decay modes are dominated by the $\rho$ and $a_1$ meson resonance respectively.  Based on the {\it number of  charged particles} in the final state, the hadronic decay modes are classified into two classes referred to as {\it 1-prong} tau decay and {\it 3-prong} tau decay.
In this work, only those $\pi$ and $\rho$/$a_1$ resonance channels which are shown explicitly in the table will be considered.
Note that those contributions comprise around 85\% of total hadronic tau decays. 

 \begin{table}[htb]
 \small
 \begin{center}
  \begin{tabular}{|c|c|c|c|c|c|c|c|} \hline
  \multirow{2}{*}{Leptonic decay} & \multicolumn{4}{|c|}{1-prong decay} & \multicolumn{2}{|c|}{3-prong decay}  \\
   & $ \pi^\pm \nu$ & $ \rho^\pm (\to \pi^\pm \pi^0) \nu$ & $a_1(\to \pi^\pm \pi^0 \pi^0) \nu$  & others & $a_1(\to \pi^\pm \pi^\pm \pi^\mp) \nu$ & others  \\ \hline
  35.2\% & 10.8\% & 25.5\% & 9.3\% & 6.4\% & 9.0 \% & 3.8\%   \\ \hline
  \end{tabular}
    \caption{\label{taubr} Tau decay channels and their branching ratios. }
    \end{center}
\end{table}

\begin{figure}[htb]
  \centering
    \includegraphics[width=0.48\textwidth]{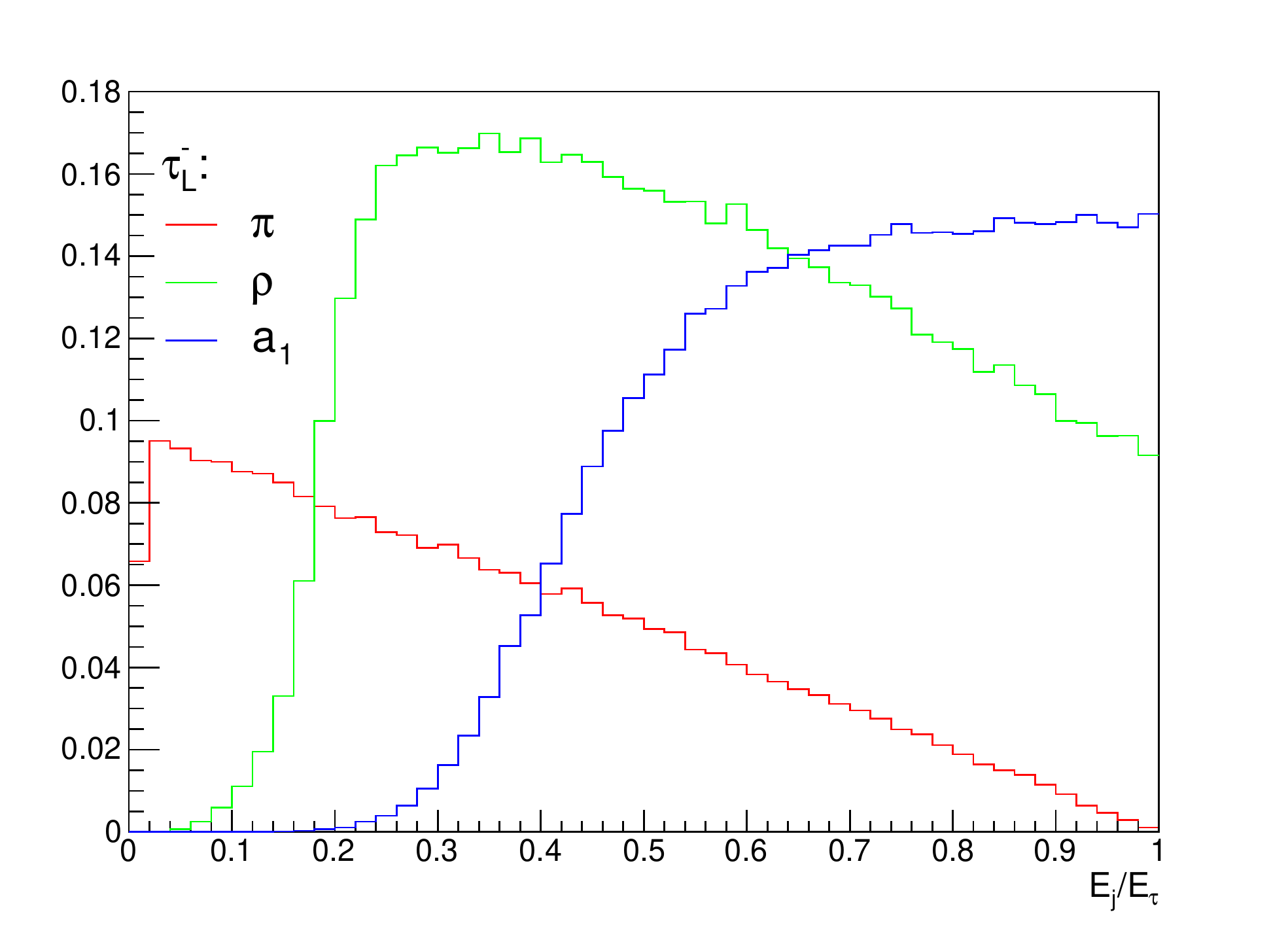}
    \includegraphics[width=0.48\textwidth]{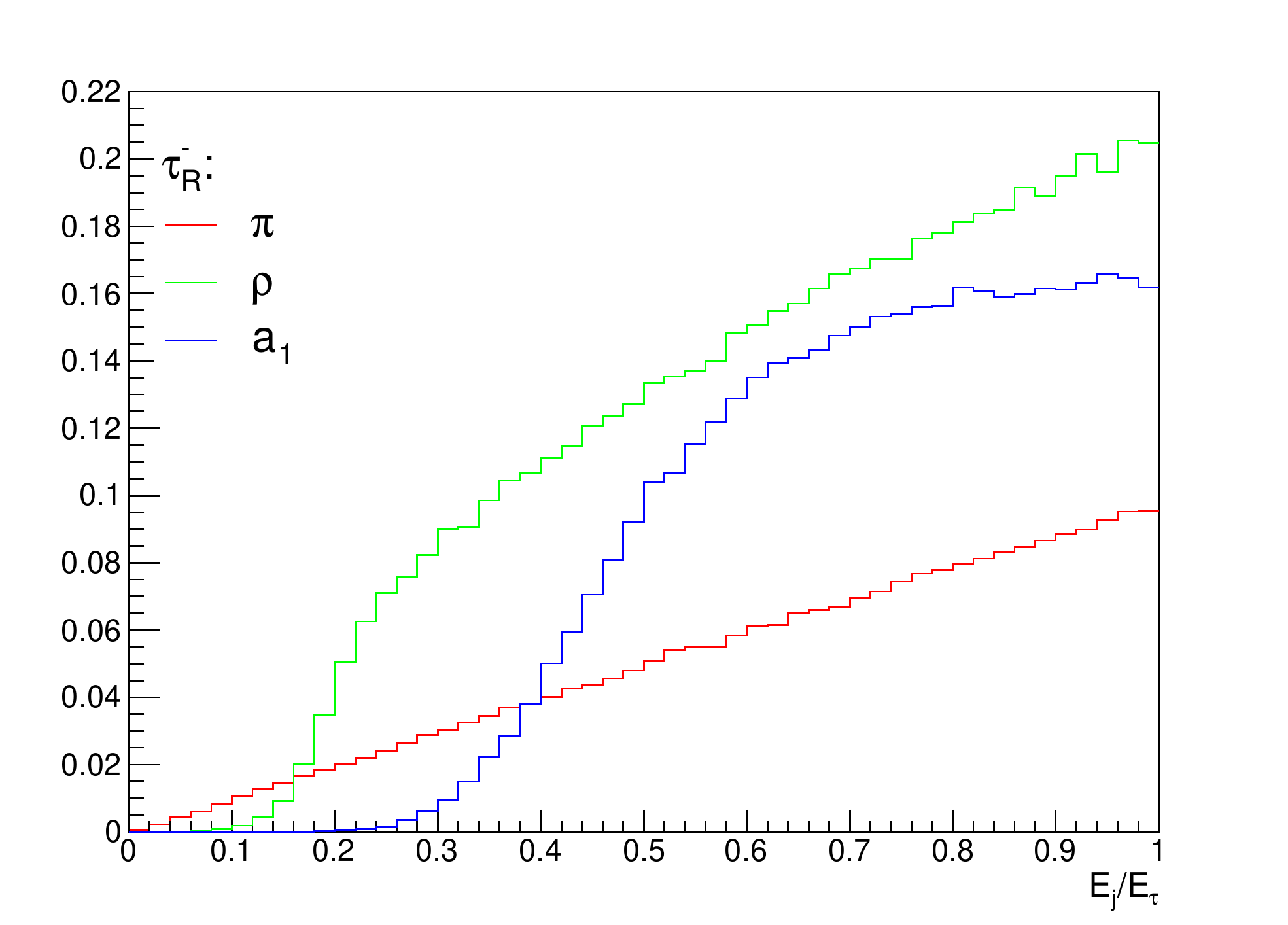} 
    \caption{\label{er} Energy ratio between visible and total final states for $\tau_{L}$ (left panel) and $\tau_{R}$ (right panel) decay. The distributions are calculated by TauDecay~\cite{Hagiwara:2012vz}. } 
\end{figure}

Since we cannot measure the momentum of the neutrino, we will not be able to fully reconstruct the tau momentum in the general case. 
However, the tau provides more information about potential new physics interactions than the first two generation of leptons through the kinematics of its decay products.  
The polarization of the tau is correlated with the energy ratio between the energy of the visible decay products and the 
energy of the tau lepton as shown in Fig.~\ref{er}.
The correlation between the energy ratio and the tau polarization depends on the decay products, where we see that the strength of
the correlation decreases as we go from $\pi$ to $\rho$ to $a_1$. However, it was shown in Ref.~\cite{Davier1993411} that by
considering more complicated multi-dimensional distributions, stronger correlations with the 
tau polarization for the $\rho$ and $a_1$ channels can be found. 
%By using that information, one can significantly improve the sensitivity
%of those channels and the $\tau \to a_1 \nu$ channel in particular. 

References~\cite{Nojiri:1996fp,Boos:2003vf,Bechtle:2009em,Schade:2009zz} were devoted to the reconstruction of the mass and the coupling of the stau based on
the $\tau \to \rho \nu$ channel. In those references the mass reconstruction was seen to suffer from limited statistics,
especially when extracting the endpoint of the  $\rho$ meson energy distribution. 
In the present work we shall try to use another remarkable feature of the tau, its relatively long lifetime, to obtain
a more precise reconstruction of new physics masses and couplings and of the spin of any new physics particle.
The tau lepton has a mean lifetime of $\tau = 2.9 \times 10^{-13}$~s and a mass of $1.777$ GeV. 
The probability of a tau traveling further than length $L$ is
$P(\beta) = e^{-L/L_0}$
with $L_0 = \gamma c \tau = \frac{c \tau}{\sqrt{1-\beta^2}} = \frac{E_\tau}{m_\tau} c\tau~.~$
So, a typical $\tau$ lepton with energy $\sim\mathcal{O}(100)$ GeV will travel a distance $l \sim 5$ mm before its decay. 
If the tau goes through a 3-prong decay, the reconstructed secondary vertex will be displaced from the primary vertex of the signal process.
From the position of the displaced vertex, we can obtain the direction of the $\tau$ lepton.
Ref.~\cite{Gripaios:2012th} showed that with the information obtained from the displaced vertex there is
an improvement for both Higgs detection and the reconstruction of the Higgs. 
A similar technique is used in this work to reconstruct the more complicated process of stau production with a subsequent decay
into a tau lepton and a neutralino. We will show that with knowledge of the tau direction, the corresponding tau energy can be reconstructed. This can be used to learn about the masses of new particles. If the directions of both tau particles in the final state
are known, the whole system can be
resolved. As a result, the angular distribution of the stau can be used to determinate its spin. 
We also propose a new method for studying the tau polarization through the impact parameter distribution of $\pi$ in the $\tau \to \pi \nu$ channel. 
This method benefits from a higher sensitivity to tau polarization~\cite{Rouge:1990kv} and is potentially more accessible experimentally, since it does not require the reconstruction of a $\rho$ candidate.

\section{New physics reconstruction}
\label{sec:npr}
In this section we consider the following  two benchmark  processes in SUSY models
that can gives rise to a $2 \tau + \slashed{E}_T$  signature. They differ in the spin of the new particles:
\begin{align}
 e^+ e^- &\to \tilde{\tau}^+_1 \tilde{\tau}^-_1 \to \tau^+ \tau^- + \tilde{\chi}^0 \tilde{\chi}^0 \to j_1 j_2 + \slashed{E}_T \label{ee2sta} \\
 e^+ e^- & \to \tilde{\chi}^+_1 \tilde{\chi}^-_1 \to \tau^+ \tau^- + \tilde{\nu} \bar{\tilde{\nu}} \to j_1 j_2 + \slashed{E}_T ~,~ \label{ee2x1}
\end{align}
where $j_{1,2}$ stand for the visible components of the $\tau$ decays and $\tilde{\chi}^0/\tilde{\nu}$ are invisible for the detector.  The chargino benchmark is studied for the purposes of making a comparison in spin reconstruction.
It will not be considered as a background for $\tilde{\tau}$ searches, since our study will be restricted to the consideration of only one new physics particle at a time.
Both processes have contributions from $s$-channel $\gamma/Z$ mediation. 
The second process may have additional contributions from a $t$-channel $\tilde{\nu}_e$ mediation, which is
suppressed in the case of using a right handed electron beam or a heavy sneutrino. 
Since a right handed electron beam is considered in our study in order to suppress the $WW$ background,
only the $s$-channel contribution needs to be considered. The same Feynman diagrams can show up in many other
new physics frameworks as well, e.g., in models with
universal extra space dimensions~\cite{Appelquist:2000nn,Cheng:2002iz}. 

In the process represented by Eq.(\ref{ee2sta}), the polarization of the produced $\tau$ is determined by the degree of mixing of the
stau sector and the neutralino sector. The bino-like neutralino gives for the tau polarization
\begin{align}
P_\tau &= \frac{B(\tau_R \tilde{\chi}^0) - B(\tau_L \tilde{\chi}^0)}{B(\tau_R \tilde{\chi}^0) + B(\tau_L \tilde{\chi}^0)} \\
& \to \left(\frac{4 \sin^2 \theta_\tau -\cos^2 \theta_\tau}{4 \sin^2 \theta_\tau +\cos^2 \theta_\tau}\right)_{ (\tilde{\chi}^0 = \tilde{B})}
\, ,
\end{align}
where $\theta_\tau$ is the mixing angle for the stau sector. 
The polarization $P_\tau$ in this case can lie in the range $[-1,1]$ when varying $\theta_\tau$. Only the
tau polarization will be of concern in our study and so the model with a bino-like $\tilde{\chi}^0$ is representative of a
large class of models with arbitrary neutralino mixing. 
We will use this model framework to study different tau polarization effects. 
The Monte Carlo events are generated using MadGraph5~\cite{Alwall:2011uj} with the TauDecay~\cite{Hagiwara:2012vz}
package used to perform simulations of the decaying tau lepton. 

\subsection{Knowledge of the tau direction from the displaced vertex}
For the two processes given above, there are two visible jets and four missing particles in the final states. At first sight
it might seem impossible to fully reconstruct the decay system. However, as mentioned in the introduction,
additional information about the tau lepton can be extracted precisely because of the relatively long tau lifetime and the
consequent displaced tau decay vertex. 

With the measurement of the $\tau$ direction from its displaced vertex, we can write the $\tau$ four-momentum as
\begin{align}
p^\mu_{\tau} &= \{ E_{\tau}, |p_{\tau}| \sin \theta \cos \phi , |p_{\tau}| \sin \theta \sin \phi, |p_{\tau}| \cos \theta \}
\end{align}
and we can measure the four-momentum for the associated jet
\begin{align}
p^\mu_{j} &= \{ E_j,  p^x_j , p^y_j , p^z_j  \}  ~,~
\end{align}
where $E_\tau = \sqrt{|p_\tau|^2 + m_\tau^2}$.
Using the fact that the neutrino is almost massless, $m_\nu^2 = (p_\tau - p_j)^2 =0$, we can then
solve the kinematics of the system up to a two fold ambiguity
\begin{align}
|p_\tau| = \frac{-2 CD \pm \sqrt{4 C^2 D^2 - 4 (D^2-4E_j^2) (C^2 - 4 E^2_j m^2_\tau)} }{2(D^2 - 4 E_j^2)}\, , 
	\label{etaumain}
\end{align} 
where
\begin{align}
C &=  m^2_\tau + m^2_j \\
D &= 2 p^x_j \sin \theta \cos \phi + 2 p^y_j \sin \theta \sin \phi + 2 p^z_j \cos \theta ~.~
\end{align}
In summary, from the jet measurements and the direction of the displaced $\tau$ vertex we can reconstruct the
$\tau$ four-momentum, $p^\mu_\tau$, up to a two-fold ambiguity.

Prior to  the $\tau$ decays our processes of interest contain two $\tau$'s and two missing neutral particles
as indicated in Eqs. (\ref{ee2sta}) and (\ref{ee2x1}).
We use the following notation in the reconstruction of the full process
\begin{align}
 p_{\tau_1} &= (E^{\tau_1}, p^{\tau_1}_x, p^{\tau_1}_y, p^{\tau_1}_z)\, , \\
 p_{n_1} &= (E^{n_1},p^{n_1}_x, p^{n_1}_y,p^{n_1}_z)\, , \\
  p_{\tau_2} &= (E^{\tau_2}, p^{\tau_2}_x, p^{\tau_2}_y, p^{\tau_2}_z)\, , \\
 p_{n_2} &= (E^{n_2},p^{n_2}_x, p^{n_2}_y,p^{n_2}_z)\ .
\end{align}

Up to the two-fold ambiguity for each of the two $\tau$ four-momentum there are eight remaining degrees of freedom.
However, we have the following eight constraints from four-momentum conservation and on-shell mass conditions in the centre
of momentum frame,
\begin{align}
 E^{n_1} + E^{\tau_1}  =  E^{n_2} + E^{\tau_2} & \equiv \sqrt{s}/2\, , \\
 \vec{p}\,^{\tau_1} + \vec{p}\,^{n_1} + \vec{p}\,^{\tau_2} + \vec{p}\,^{n_2} & = 0\, , \\
 (E^{n_1})^2-(p^{n_1}_x)^2 - (p^{n_1}_y)^2 - (p^{n_1}_z)^2 &= m^2_{n}\, , \\
  (E^{n_2})^2- (p^{n_2}_x)^2 - (p^{n_2}_y)^2 - (p^{n_2}_z)^2 &= m^2_{n}\, , \\
  (E^{\tau_1}+E^{n_1})^2-(p^{\tau_1}_x-p^{n_1}_x)^2 - (p^{\tau_1}_y+p^{n_1}_y)^2 - (p^{\tau_1}_z+p^{n_1}_z)^2 &= m^2_{\tilde{\tau}} \, 
\end{align}

By solving this system of equations we find
\begin{align}
E^{n_i} &= \frac{\sqrt{s}}{2} - E^{\tau_i}, ~ i=1,2  \\
\vec{p}\,^{n_2} &= -\vec{p}\,^{\tau_1} - \vec{p}\,^{\tau_2} - \vec{p}\,^{n_1} \\
p^{n_1}_x &= D_1 + D_2 p^{n_1}_y +D_3 p^{n_1}_z  \\
p^{n_1}_y &= F_1 + F_2 p^{n_1}_z  \\
p^{n_1}_z &= \frac{-(G_1G_2+F_1F_2) }{G_2^2 + F_2^2 +1} \\
                 & \pm \frac{\sqrt{(G_1 G_2 + F_1 F_2)^2 - (G_2^2+ F_2^2 +1)(m_n^2+G_1^2+F_1^2-(E^{n_1})^2)} }
                 		{G_2^2 + F_2^2 +1} \, ,
\end{align}
where
\begin{align}
C_1 & = (E^{n_2})^2-(E^{n_1})^2 - (\vec{p}\,^{\tau_1+\tau_2})^2, \\
D_1 & = \frac{C_1}{2p_x^{\tau_1+\tau_2}} ,  ~
D_2  = -\frac{p_y^{\tau_1+\tau_2}}{p_x^{\tau_1+\tau_2}} ,  ~
D_3  = -\frac{p_z^{\tau_1+\tau_2}}{p_x^{\tau_1+\tau_2}} ,\\
E_1 &= 2 E^{n_1} E^{\tau_1} -2 p_x^{\tau_1} D_1 + (m^{n_1})^2+ (m^{\tau})^2 - (m^{\tilde{\tau}})^2, \\
F_1 & = \frac{E_1}{2(p^{\tau_1}_x D_2 + p_y^{\tau_1})},\\
F_2 & = -\frac{p_x^{\tau_1} D_3 + p_z^{\tau_1}}{p^{\tau_1}_x D_2 + p_y^{\tau_1} },\\
G_1 & = D_1 + D_2 F_1 ,\\
G_2 & = D_2 F_2 +D_3 \, .
\end{align}
There is a two fold ambiguity also for this system of equations.

By reconstructing the kinematics in this manner, we will 
have on overall eight-fold ambiguity made up as $2(\text{from }\tau_1)\times 2(\text{from }\tau_2)\times 2(\text{from final reconstruction})=8$.
We will next show that the new particles mass and spin can be reconstructed by using this method.

\subsubsection{Reconstructing new particle masses}
\label{pmass}
For the  processes described in Eqs.~(\ref{ee2sta}) and (\ref{ee2x1}) the $\tau$ energy distribution is bounded from  above and below with its end points given by
\begin{align}
E^\tau_{\max(\min)} = \frac{E^*_\tau \pm p^*_\tau \beta_{\tilde{\tau}}}{\sqrt{1-\beta^2_{\tilde{\tau}}}} \, , \label{emm}
\end{align}
where
\begin{align}
E^*_\tau = \frac{m^2_{\tilde{\tau}}-m^2_{\tilde{\chi}^0_1} + m^2_\tau }{2 m_{\tilde{\tau}}}, ~~ p^*_{\tau} = \sqrt{(E^*_\tau)^2 - m^2_\tau}, ~~ \beta_{\tilde{\tau}}= (1-4 m^2_{\tilde{\tau}}/s)^{1/2} ~.~ \label{emm_2}
\end{align} 
By studying the energy distribution of the tau lepton, we will be able to reconstruct the mother particle ($\tilde{\tau}$ or
$\tilde{\chi}^\pm$) mass and missing particle ($\tilde{\chi}^0$ or $\tilde{\nu}$) mass. 

\begin{figure}[htb]
  \centering
    \includegraphics[width=0.48\textwidth]{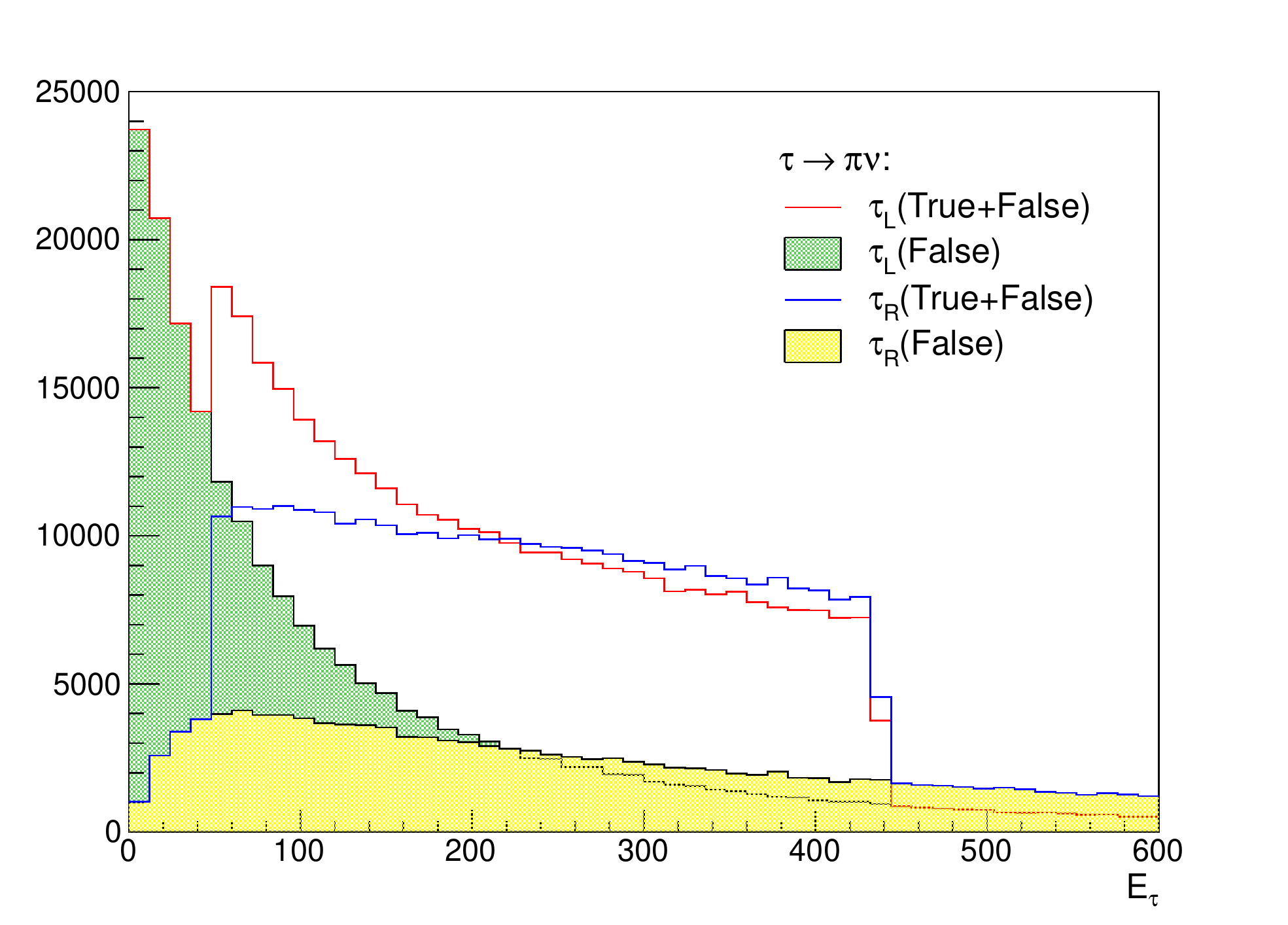} 
    \includegraphics[width=0.48\textwidth]{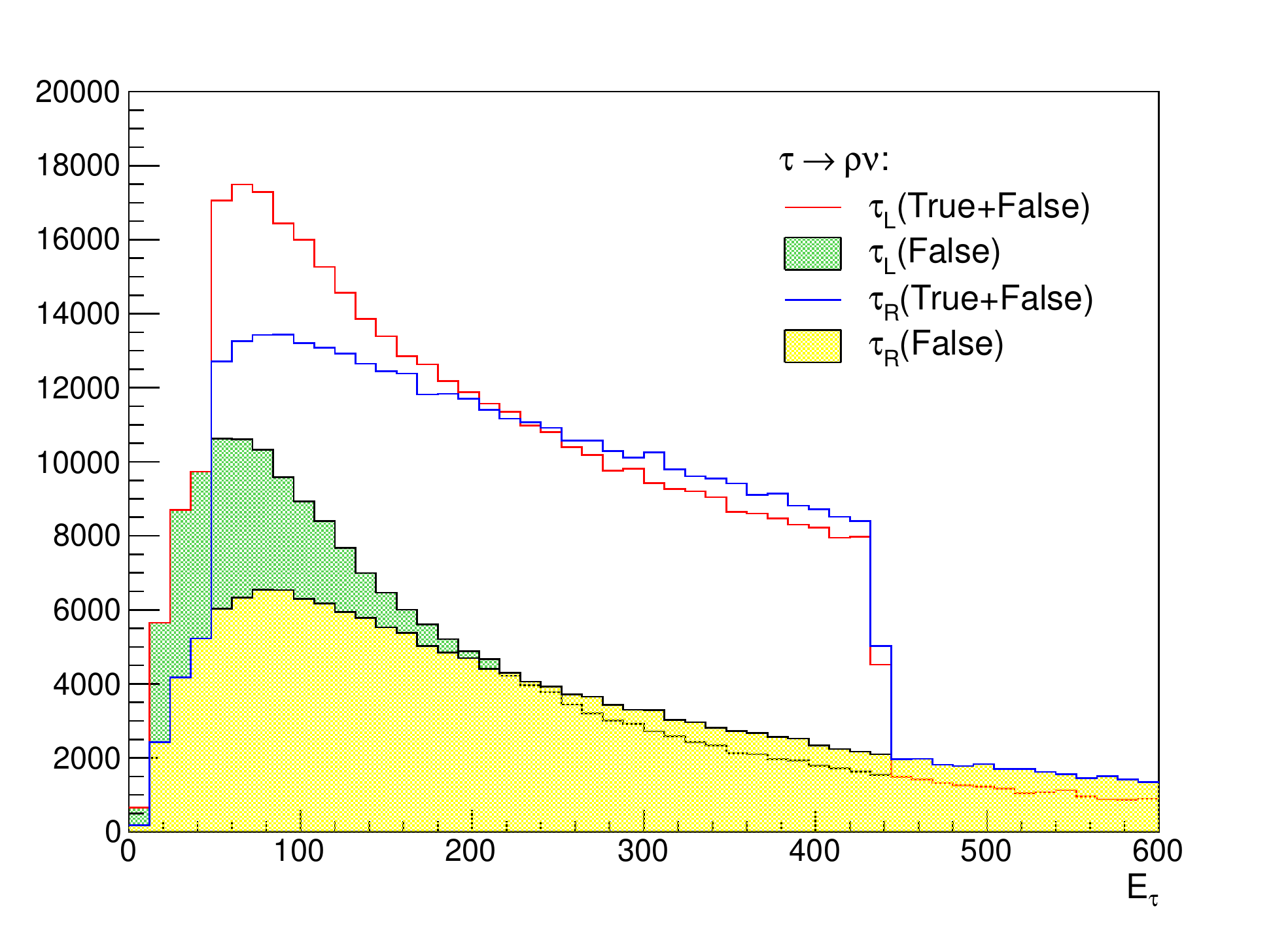} \\
        \includegraphics[width=0.48\textwidth]{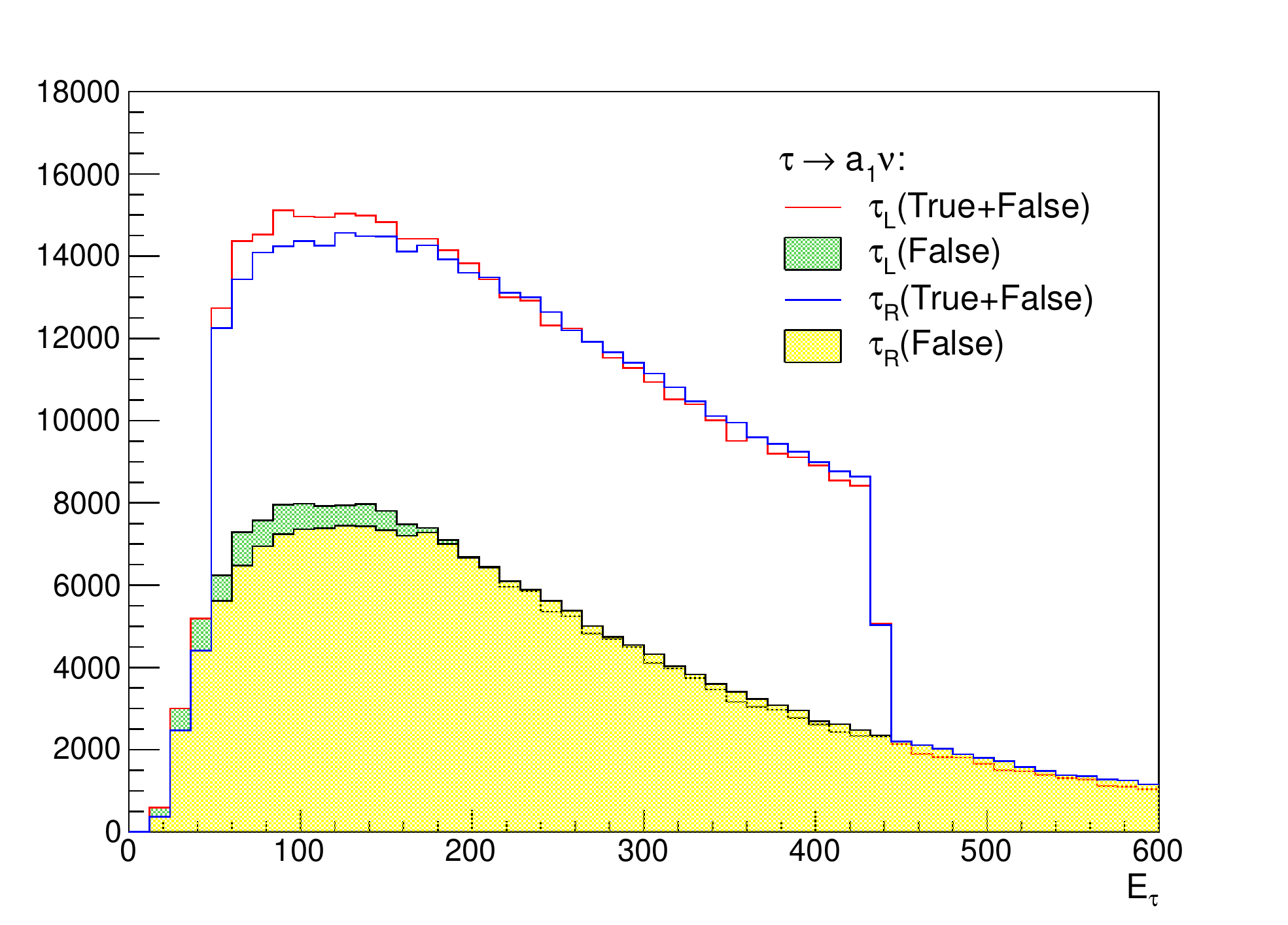}
          \includegraphics[width=0.48\textwidth]{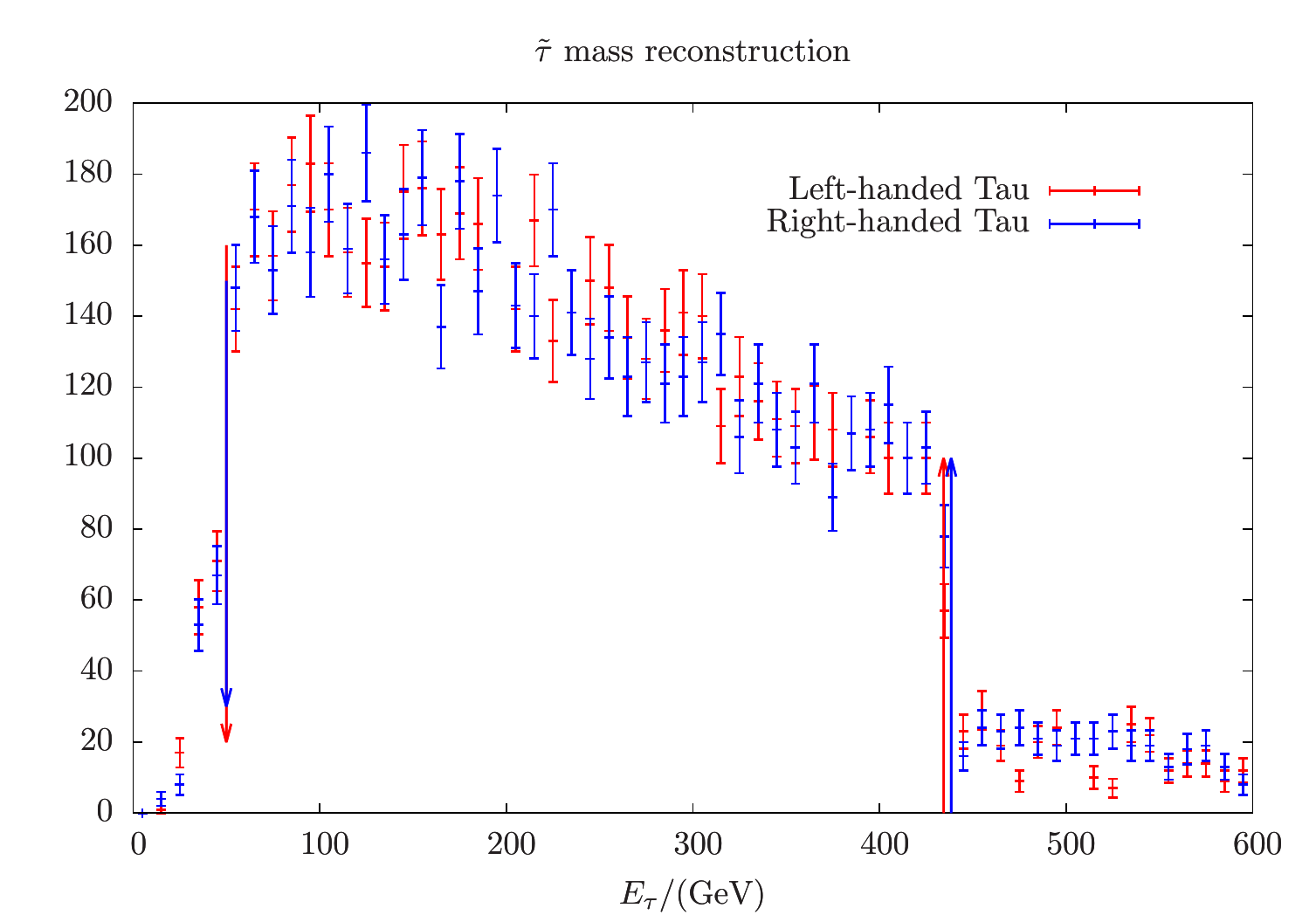}   
    \caption{\label{etau} Reconstructed $\tau$ energy for different chiralities and decay topologies: $\tau \to \pi \nu$ (upper left), $\tau \to \rho \nu$ (upper right) and $\tau \to a_1 \nu$ (lower left). The lower right panel gives the full tau energy distribution for $m_{\tilde{\tau}} =300$ GeV and $m_{\tilde{\chi}^0}=50$ GeV. The arrowed lines show the reconstructed endpoints ($E_{\max/\min}$) for $E_{\rm true}$ distributions. }
\end{figure}

With the four momentum of the visible sector from the $\tau$ decay and the direction of the $\tau$ momentum, the energy of the
$\tau$ can be reconstructed up to a two-fold ambiguity.  Up to small changes due to next-to-leading order (NLO) corrections and detector effects, we know from two body decay kinematics~\cite{Tsukamoto:1993gt} that the distribution of the ``true" tau energy solution is flat over the entire allowed energy range, $E_{\min}\le E \le E_{\max}$, and independent of the tau polarization.
In general we expect the energy distribution of the $\tau$ decay products to depends on the tau polarization, which may lead to different distributions of the false $E_\tau$ solution for different tau polarization. By considering the $\tilde{\tau}$ pair production process 
we show the distribution of $E_{\rm false}$ and $E_{\rm all}$ for three different tau decay channels in Fig~\ref{etau}, where we have taken $m_{\tilde{\tau}} =300$ GeV, $m_{\tilde{\chi}^0}=50$ GeV and $\sqrt{s}=1$ TeV.

From the figure, we can conclude that for the $\tau \to \pi \nu$ and the $\tau \to \rho \nu$ decay channels,
the false tau energy solution distributions are strongly dependent on the tau polarization.
As a result, it will be difficult to extract the true tau energy distribution without knowing the tau polarization. The situation is different for the $\tau \to a_1 \nu$, which gives the dominant contribution to 3-prong tau decay. 
For this channel, different tau polarizations give almost identical false tau energy distributions. Moreover, the tau direction in this case can be reconstructed from the  location of the secondary vertex.

For illustration purposes, a sample of 3000 three-prong tau decays is used to study the mass reconstruction precision.
The superposed distribution of $E_{\rm true}$ and  $E_{\rm false}$ are shown in the lower right panel of Fig.~\ref{etau}  with statistical uncertainty built into the distribution. 
The full (true+false) distribution is comprised of a rectangular $E_{\rm true}$ distribution and a continuous $E_{\rm false}$ distribution. We use following algorithm to locate the endpoints of the $E_{\rm true}$ distribution from the full distribution: 
\begin{itemize}
\item The location of the falling edge is estimated to be where $(N_{i-1}-N_{i+1})/\sqrt{N_{i}}$ is maximized, where $N_{i}$ is number of events in $i$th bin;
\item The height of the rising edge which is given by the height of the rectangular $E_{\text{true}}$ distribution, can be estimated by  $h\equiv N_{i-1}-N_{i+1}$;
\item Allowing some fluctuations, the largest $j$ such that $N_{j}-N_{j-1} > 0.8 h$ can be used to locate the rising edge. Note only two adjacent bins are used here because of the better statistics due to the shape of the false distribution; 
\item Improved estimates of the location of the edges are given by by $E_{\max}= E_{i} + \frac{S}{2} \frac{2N_i  - N_{i+1} -N_{i-1} }{N_{i-1}-N_{i+1}}$ and $E_{\min}= \frac{E_{j}+E_{j-1}}{2}$, where $E_{i}$ is the center value of the $i$th bin and $S$ is the bin size; 
\item The corresponding uncertainties are $\delta E_{\min} = S/\sqrt{8}$ and 
\begin{align}
\delta E_{\max} = \sqrt{(S/\sqrt{12})^2+ \sum_{k=i-1,i,i+1} (\frac{\partial E_{\max}}{\partial N_k} \sqrt{N_k})^2 }. 
\end{align}
\end{itemize}

The reconstructed central values of $E_{\max}$ and $E_{\min}$ are shown by arrowed lines in the lower right panel of Fig.~\ref{etau}, for the left-handed tau ($E_{\min}=50.0\pm 3.5$ GeV, $E_{\max}=434.4 \pm 3.2$ GeV) and the right-handed tau ($E_{\min}=50 \pm 3.5$ GeV, $E_{\max}=438.5\pm 3.5$ GeV) respectively.
The new particles masses can then be reconstructed by using Eqs.~(\ref{emm}) and (\ref{emm_2}). For left-handed tau we
obtain $m_{\tilde{\tau}}=304.2^{+9}_{-10} $ GeV, $m_{\tilde{\chi}^0}=53.8^{+6}_{-8}$ GeV and for right-handed tau we
find $m_{\tilde{\tau}}=303.1^{+9}_{-9}$ GeV, $m_{\tilde{\chi}^0}= 46.0^{+7}_{-9}$ GeV, which compare well with our input values of  $m_{\tilde{\tau}} =300$ GeV, $m_{\tilde{\chi}^0}=50$ GeV.

\subsubsection{Reconstructing new particle spin}
For the $s$-channel $\gamma/Z$ mediated scalar/fermion production at an
$e^+ e^-$ collider, the spin of the new particles can be related to
threshold excitation and angular distribution~\cite{Choi:2006mr}. 
Because of our relatively heavy benchmark point, accumulating a sufficient number of events at different collision energies 
would require a large amount of accelerator operation time. 
Therefore, we will focus exclusively on the angular distribution as a spin discriminator in this work. 
The polar angle distribution for the scalar pair production in the $s$-channel in the
Centre of Mass Frame (CMF) is
\begin{align}
\frac{1}{\sigma} \frac{d \sigma}{ d \cos \theta_{\tilde{\tau}}} [e^+ e^- \to \tilde{\tau} \tilde{\tau}]  \propto \sin^2 \theta_{\tilde{\tau}} = 1- \cos^2 \theta_{\tilde{\tau}} \, , \label{CMF_sigma_scalar}
\end{align}
whereas for fermion pair production in the $s$-channel we have
\begin{align}
\frac{1}{\sigma} \frac{d \sigma}{ d \cos \theta_{\tilde{\chi}^\pm}} [e^+ e^- \to \tilde{\chi}^\pm \tilde{\chi}^\mp] \propto  (s-4 m^2_{\tilde{\chi}^\pm}) \cos^2  \theta_{\tilde{\chi}^\pm} + (s + 4 m^2_{\tilde{\chi}^\pm}) ~.~\label{CMF_sigma_fermion}
\end{align}
From these equations, we can see that a scalar particle tends to be produced in the central region, 
whereas the fermion is more likely to be produced in the forward/backward regions. 
It can be shown that the $t$-channel $\tilde{\nu}$ mediated process for fermion final states will 
lead to an asymmetric distribution for $\cos \theta_{\tilde{\chi}^\pm}$ when only one particular charge is considered~\cite{Zhu:2004ei}. 
For each electric charge of the final state, either the forward or the backward direction is favored.
In particular, when the sneutrino mediator is light,  the final state is much more concentrated in the forward/backward regions
than the $s$-channel process, which helps make the spin characteristic even more distinguishable.  
However, the $t$-channel process is highly suppressed for a right-hand polarized electron beam and may even be non-existent
in some new physics scenarios. For this reason, we do not consider this subprocess in our discussions. 

Using Eqs.~(\ref{CMF_sigma_scalar}) and (\ref{CMF_sigma_fermion}), we can reconstruct the polar angle distribution for $\tilde{\tau}/\tilde{\chi}^\pm$ in order to extract spin information about the new particles.
As we have noted before, our method typically produces an 8-fold ambiguity for our system with only one of them being physical.  The experimentally accessible variable is the 8-fold superposition of all solutions.  
In practice, the distribution of the seven false solutions will inherit information about the true physical solutions and will be
different for scalar and fermion particles. 
We will restrict our considerations to the case where there is only one new physics process at a time. For example, we do not consider here
the possibility of a superposition of the effects of both scalar and fermion new physics particles. 

\begin{figure}[htb]
  \centering
    \includegraphics[width=0.48\textwidth]{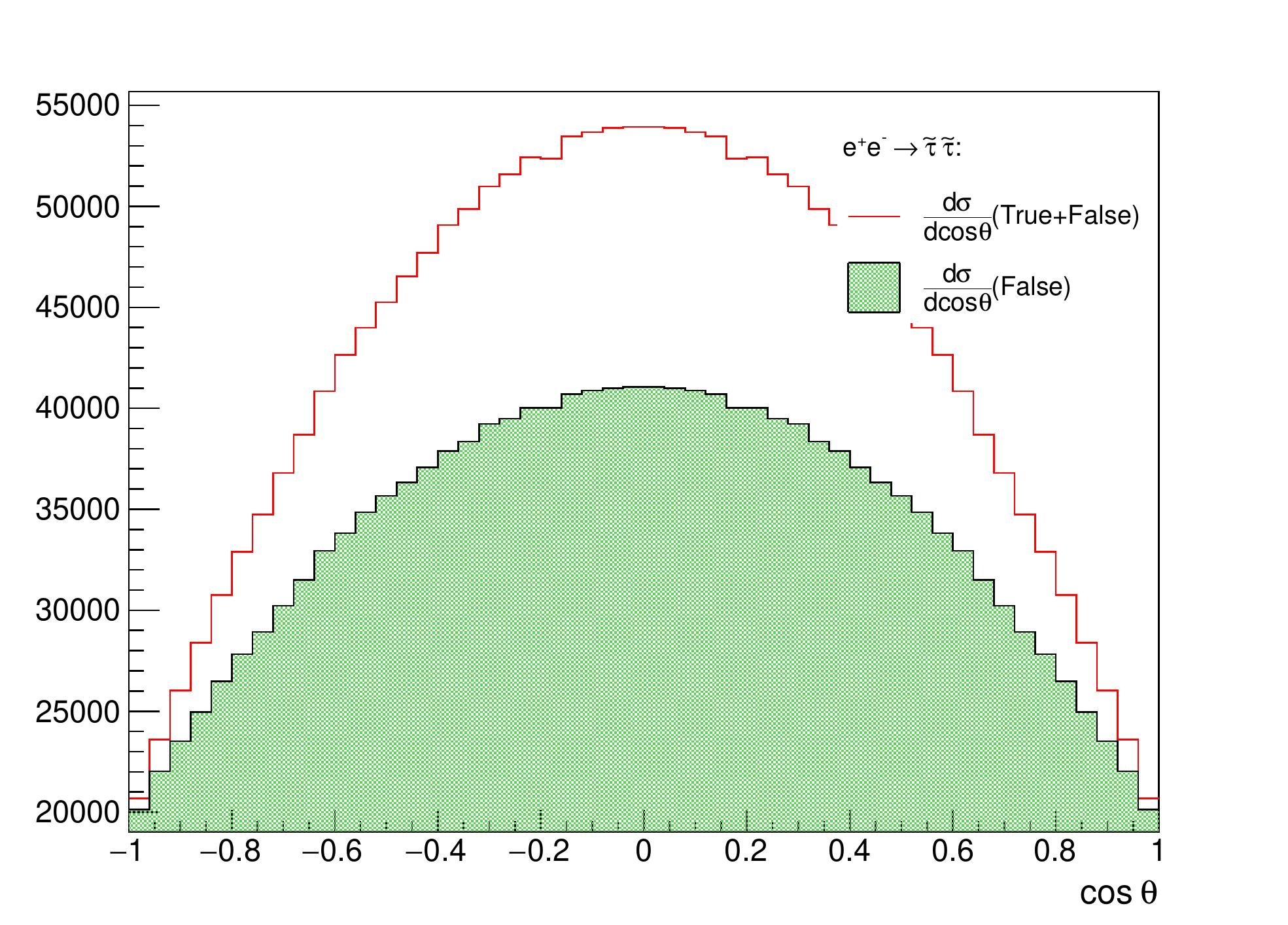} 
    \includegraphics[width=0.48\textwidth]{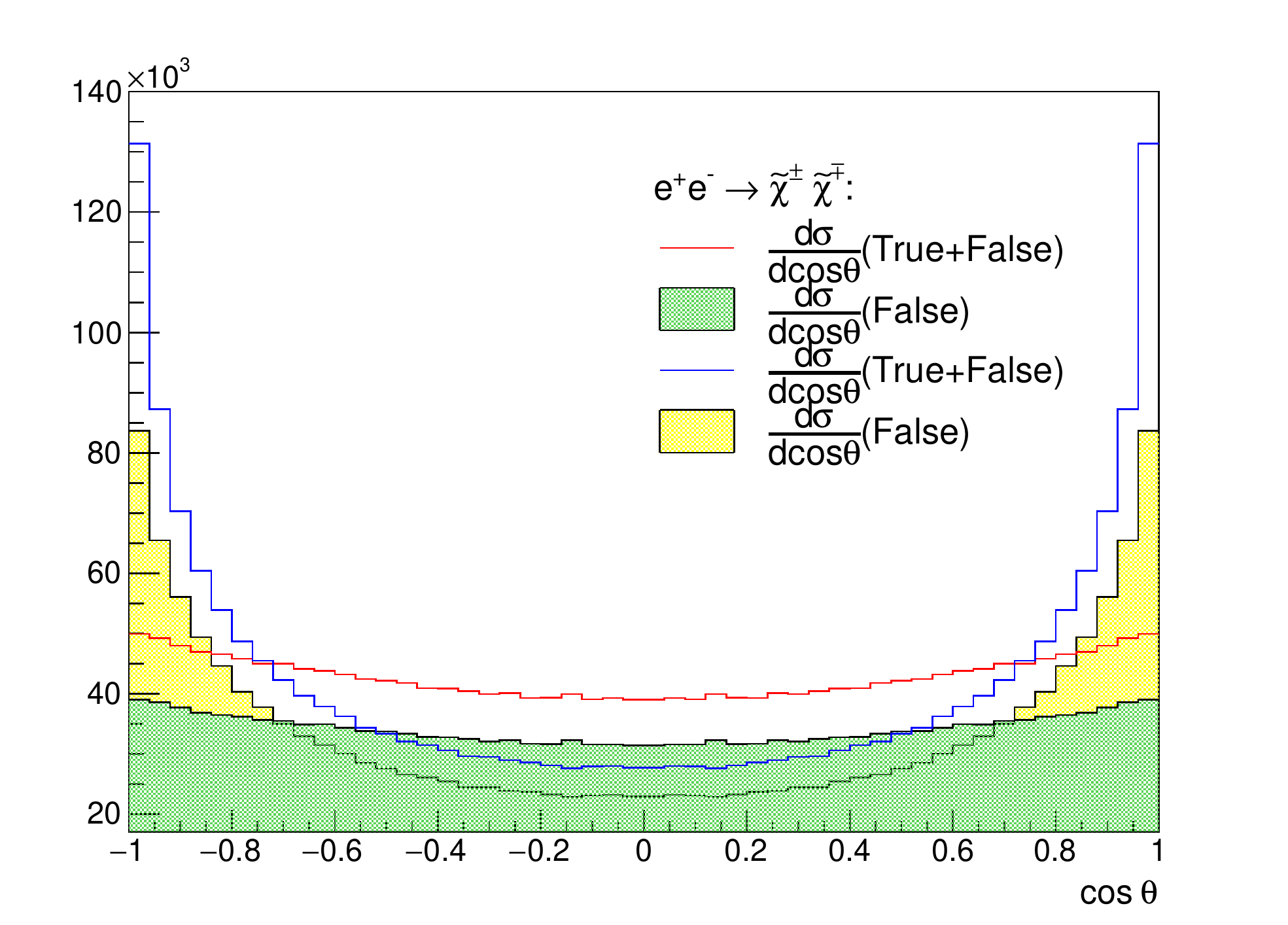} 
    \caption{\label{spin} The $\cos \theta_{\tau}$ distribution for $\tilde{\tau}$ (left panel) and chargino (right panel) pair production at a 1 TeV $e^+ e^-$ collider.  The red line and green shadowed region in both plots correspond respectively to the True+False and the False solution distributions for the $s$-channel process. The blue line and yellow shadowed region in the right plot are respectively the True+False and the False solution distributions for the $t$-channel fermion production.  }
\end{figure}

The polar angular distribution of the sum of the 7 false solutions (denoted as False) and all 8 solutions (denoted as True+False) for $\tilde{\tau}$ and $\tilde{\chi}^\pm$ are shown in Fig.~\ref{spin}. In the figure we include both charges of the produced particles ($\tilde{\tau}$, $\tilde{\chi}^\pm$), which will lead to a symmetric distribution  for the $t$-channel process.  From the figure we find for all cases that the corresponding false solutions are distributed similarly but are flatter than the true solution. The $t$-channel contribution is shown as well, where the sneutrino mass is taken as 100 GeV.  It is featured by a more remarkable concentration at forward and backward region than $s$-channel fermionic final state process. 

In order to measure the difference between particles spins, we  can define a spin measure on the superposed distribution of all eight solutions (corresponding to True+False):
\begin{align}
A_s =\frac{N(|\cos(\theta)|<0.5) -N(|\cos(\theta)|>0.5)}{N(|\cos(\theta)|<0.5) + N(|\cos(\theta)|>0.5)}~.~
\end{align}
We see from Fig.~\ref{spin} that $A_s$ is positive for scalars and negative for fermions.
As discussed above, we will only consider the $s$-channel process here.
Using a very large number of simulated events we construct the red lines in Fig~\ref{spin} to high precision. We can then
extract $A_s$. We refer to this as the theoretical  prediction for the spin measure and denote it as $A^{\text{th}}$. We find  $A^{\rm th}_{\text{fermion}}= -0.065$ and $A^{\rm th}_{\text{scalar}}= 0.2$. 
The corresponding statistical uncertainty~\cite{Suehara:2009nj} can be estimated by
\begin{align}
\delta A_s =\sqrt{ \left(\frac{\partial A_s}{\partial N_<} \delta N_< \right)^2+\left( \frac{\partial A_s}{\partial N_>} \delta N_> \right)^2} = \sqrt{\frac{1-A_s^2}{N}},  \label{staerr}
\end{align}
where $N_<$ and $N_>$ stands for $N(|\cos(\theta)|<0.5)$ and $N(|\cos(\theta)|>0.5$) respectively. The statistical uncertainty of $N_{<(>)}$ is $\delta N_{<(>)} = \sqrt{N_{<(>)}}$.
Because there is a correlation between those 8-fold solutions, the total number of events $N$ is used to give a conservative estimate for the statistical uncertainty rather than $8\times N$.
As a result, without taking into account any detector effects, 
an event sample number of
\begin{align}
N \sim 9 \times \frac{1-(A^{\rm th}_{\text{scalar}})^2}{(A^{\rm th}_{\rm scalar} - A^{\rm th}_{\rm fermion})^2} \sim 120
\end{align}
is sufficient to distinguish the stau spin at the 3$\sigma$ level.

\subsection{Tau polarization from impact parameter}
Tau polarization can be inferred in all decay channels in different ways. The  simplest case is 
$\tau \to \pi \nu$ with the left handed interaction~\cite{Hagiwara:2012vz}
\begin{align}
\mathcal{L} = \sqrt{2} C \bar{\tau} \gamma^\mu P_L \nu_\tau \partial_\mu \pi  + h.c. ~.~
\end{align}
In the tau rest frame, the relation between tau polarization($P_\tau$) and the pion angular distribution($\cos \theta$) is given by,
\begin{align}\label{Eq:dGamma/dcostheta}
\frac{1}{\Gamma} \frac{d \Gamma}{ d \cos \theta} =\frac{1}{2} (1+ P_\tau \cos \theta).
\end{align}
After boosting the whole system to the Laboratory frame (LAB), we will have 
\begin{align}
E_\pi &= \gamma \sqrt{p^2_\pi + m^2_\pi} + \beta \gamma p_\pi \cos \theta ~,~\\
E_\tau &= \gamma m_\tau ~,~
\end{align}
where $\gamma$ is the boost and where $p_\pi = (m^2_\tau - m^2_\pi)/2 m_\tau$ and $\theta$ are the magnitude of the pion three-momentum and its polar angle in the tau decay rest frame respectively. 
The energy fraction of the pion in the LAB frame $x= E_\pi/E_{\tau}$ is linearly related to the pion polar angle in the tau rest frame 
for a highly relativistic tau ($\beta\to 1$), since we have
\begin{align}\label{x_defn}
x = \frac{E_\pi}{E\tau} = (1+\frac{m^2_\pi}{m^2_\tau})\frac{1}{2} + \beta(1-\frac{m^2_\pi}{m^2_\tau})\frac{\cos \theta}{2}
\overset{\beta\to 1}{\longrightarrow}(1+\frac{m^2_\pi}{m^2_\tau})\frac{1}{2} + (1-\frac{m^2_\pi}{m^2_\tau})\frac{\cos \theta}{2} ~.~
\end{align}
So, from Eq.~(\ref{Eq:dGamma/dcostheta}) we find in the Lab frame after a change of variable from $\cos\theta$ to $x$
\begin{align}
\frac{1}{\Gamma} \frac{d \Gamma}{d x} & = \frac{1}{1- (m_\pi / m_\tau)^2} + P_\tau \frac{2 x - (1+(m_\pi / m_\tau)^2)}{(1- (m_\pi / m_\tau)^2)^2} \\ 
 &\sim 1+ P_\tau (2 x -1) \, ,
\end{align}
which can be used for studying tau polarization~\cite{Hagiwara:1989fn}.  

However, in the $\pi$-channel, the energy of $\tau$ lepton cannot be easily reconstructed for most processes of interest, since there is usually more than one missing particle in the final state. Exceptions arise in some very special cases, e.g., (i) for a few TeV tau lepton whose energy can be measured by its track curvature directly, (ii) for single tau production with a hadronic decay, and (iii) for tau pair production at electron-positron colliders. Those exceptions are beyond the scope of our current study.  The $\pi$ energy spectra can also be used directly to measure the tau polarization for known new physics process~\cite{Hagiwara:1989fn,Nojiri:274380,Schade:2009zz}. However, the $\pi$ energy distribution depends on the process under consideration and on the masses of new physics particles.

In most studies~\cite{Boos:2003vf,Bechtle:2009em,Berggren:2015qua} at colliders the $\tau \to \rho(\to \pi^+ \pi^0) \nu$ channel is chosen for measuring the tau polarization due to the fact that a $\tau^-_R$ decays mostly to a longitudinally polarised $\rho$ meson while a $\tau^-_L$ decays  mostly to a transversely polarised $\rho$. The energy ratio $z =E_{\pi^\pm} / E_{\rho}$ can be used to measure the polarization of  the $\rho$ meson, which in turn gives information about the tau polarization. The distribution of $z$ with respect to the tau polarization is given in Ref.~\cite{Hagiwara:1989fn}. 
Their equations show that the $z$ distribution is related to the tau polarization in a complicated way. Studies in Ref.~\cite{Rouge:1990kv} have given the general relative polarization sensitivity for each tau decay channel. They found that the $\tau \to \rho \nu$ channel does not perform as well as the $\tau \to \pi \nu$ channel, while the three prong decay and leptonic decay channels only have 1/3 the sensitivity of the $\tau \to \pi \nu$ channel. Another difficulty arises for the $\rho$ channel in practice. One needs to have an efficient tagger on the $\rho$ meson in order to recognise the decay topology since both mis-tagged tau jets and other 1-prong tau decay modes have to be excluded first. 

The proposed ILC has the advantage that it can resolve the track precisely. As mentioned in the introduction, an impact parameter resolution as small as 5 $\mu \text{m}$ can be reached in the $r\phi$ plane. This will be very helpful if we try to study the tau polarization using the $\tau \to \pi \nu$ channel. For those  processes of interest, the tau lepton energy has a flat distribution between $E_{\min}$ and $E_{\max}$.  Moreover, different tau polarizations give rise to very different distributions of the pion polar angle in the tau rest frame. This means that in the laboratory frame we will have a different impact parameter distribution for a different tau polarization, which in turn is related to the angle difference  between the tau and pion momenta, $\theta_{\pi - \tau}$.
The impact parameter is given by
\begin{align}
d = L \sin \theta_{\pi - \tau} \, ,
\end{align}
where $L$ is tau decay length. Therefore the impact parameter is dependent on the tau polarization. 

A difficulty arises in a realistic detector. 
Because of the long lifetime of the tau, there are no long-lived hard tracks associated with the interaction point (IP). In addition, the interaction region is relatively wide along the z-axis Therefore we can not have good resolution for the position of IP. 
 As a result, the impact parameter becomes experimentally unreconstructable. Thanks to the narrow beam size along the vertical and horizontal directions, the position of  IP can still be measured precisely in the $r\phi$ plane.  
Then, we can define the geometrically signed impact parameter~\cite{Abreu:1993uc} which is the distance of closest approach of the extrapolated track to the assumed production point, the centre of the interaction region, in the $r\phi$ plane,
\begin{align}
d_{\text{GIP}} = L \sin \theta_\tau \sin (\phi - \phi_\tau) \, ~.~ \label{eqgip}
\end{align}

\begin{figure}[htb]
  \centering
    \includegraphics[width=0.48\textwidth]{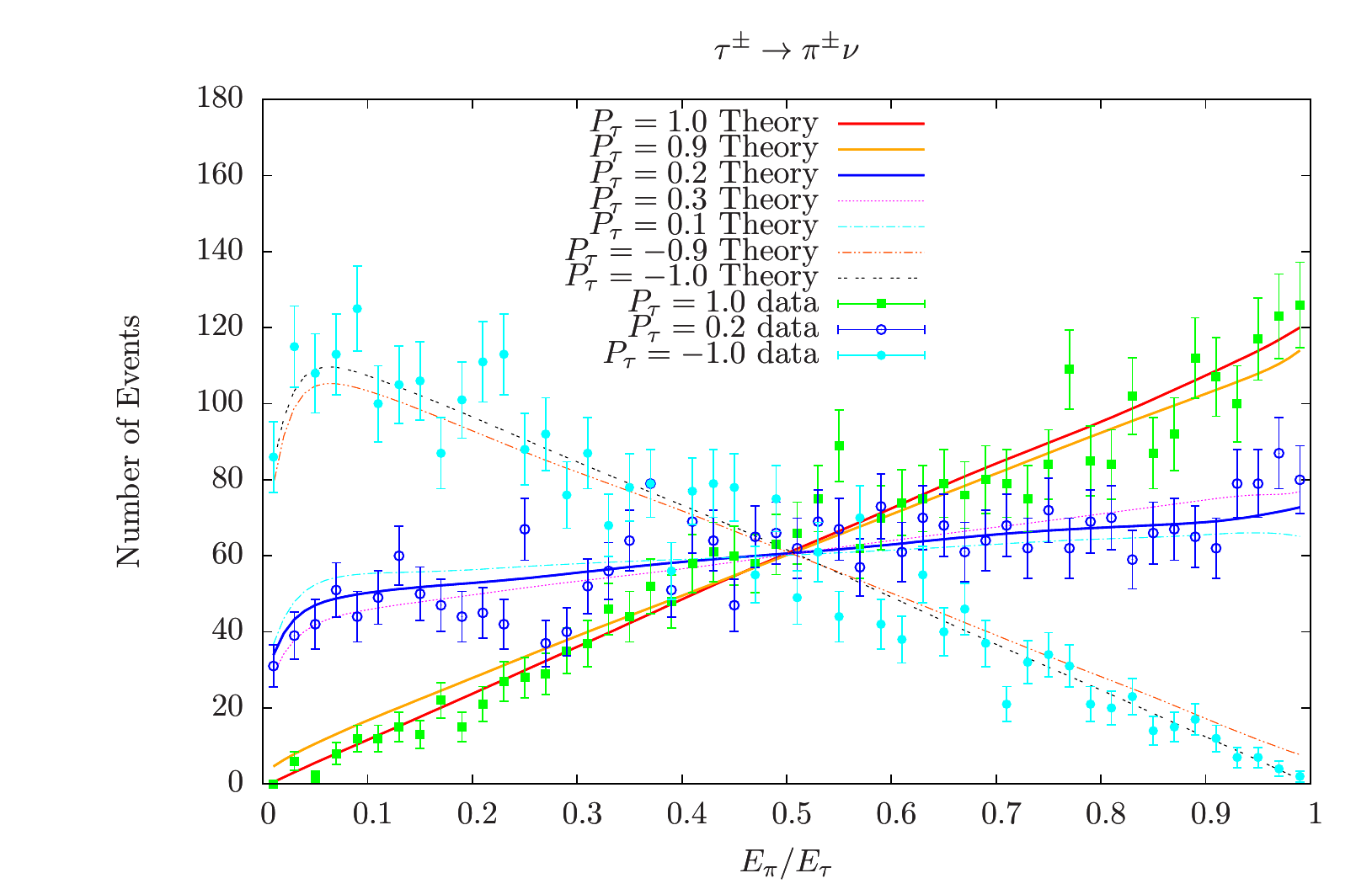} 
    \includegraphics[width=0.48\textwidth]{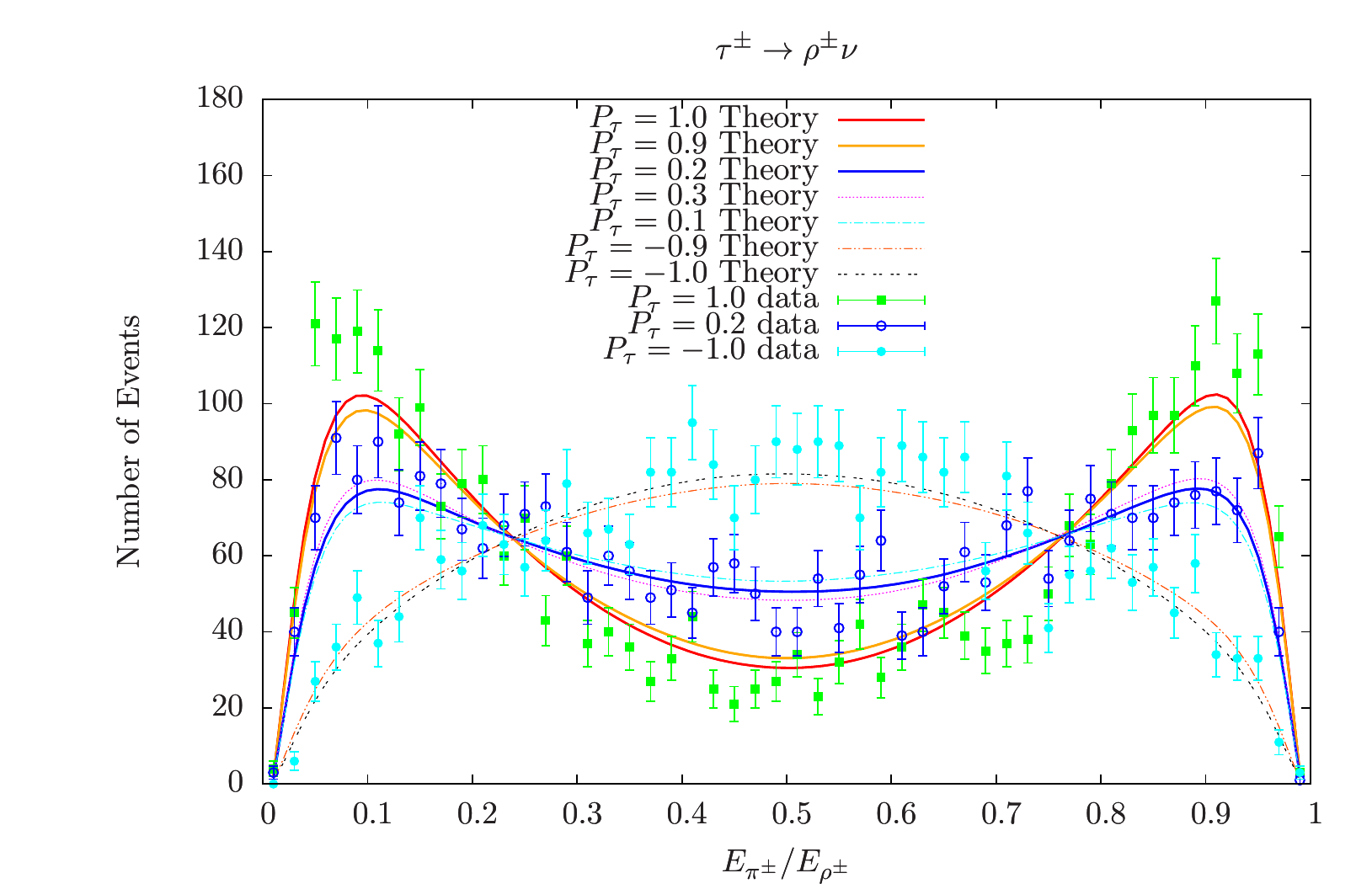} 
    \caption{\label{polfit} Theoretical and MC truth distributions by using discriminate variables constructed from $\tau \to \pi \nu$ (left panel) and $\tau \to \rho \nu$ (right panel).  
   The MC truth distribution is using 3000 tau decays with  statistical uncertainties of $\sqrt{N_i}$ for each bin $i$ .} 
\end{figure}

In the following, we shall study the sensitivities of variables to the tau polarization. For the $\tau \to \pi \nu$ channel, we have shown in Eq.~(\ref{x_defn}) that the energy ratio ($x = E_\pi / E_\tau$) in the LAB frame has the same sensitivity as the polar angle $\theta$, since they are linear related with each other in the collinear (i.e., relativistic) limit.  
It should be noted that neither $x$ nor $\theta$ are reconstructable experimentally for the stau pair production process,
except in special circumstances as we noted earlier. However, we will use $x$ extracted from the Monte Carlo data as ``best-case reference" for the purposes of comparison with the impact parameter $d_{\text{GIP}}$.
We first study the $x = E_\pi / E_\tau$ distribution for different tau polarizations, focusing on three benchmark points, $P_\tau=1.0$, $P_\tau = 0.2$ and $P_\tau =-1.0$.  We use these to demonstrate the sensitivity to  right-handed, highly mixed and left-handed polarized tau, respectively. The variation of the polarization by $\sim$ 0.1 around these three benchmark points is also studied for comparison. The $x$ distributions are given in the left panel of Fig.~\ref{polfit} for all those 7 cases. In addition, the Monte Carlo data with limited statistics (3000 tau decays) for those three benchmark points are superimposed on the plot. The error bars show the corresponding statistical uncertainties. 
To have an intuitive comparison with other variables we define the following measure for tau polarization,
\begin{align}
A_{R\pi}= \frac{N_{x<0.5} - N_{x>0.5}}{N_{x<0.5} + N_{x>0.5}} ~,~
\end{align}
which is sensitive to the shape of the dependence on $x$.
As for the $\tau \to \rho \nu$ channel, we show the energy ratio of $z=E_{\pi^\pm} / E_\rho$ in the right panel of Fig.~\ref{polfit}, where a similar analysis to that used for $\pi \nu$ channel has been used. From the figure, we can define another measure of the polarization sensitivity for $z$,
\begin{align}
A_{R\rho} = \frac{N_{|z-0.5|<0.25} - N_{|z-0.5| > 0.25}}{N_{|z-0.5|<0.25} + N_{|z-0.5| > 0.25}} 
\end{align} 

\begin{figure}[htb]
  \centering
     \includegraphics[width=0.48\textwidth]{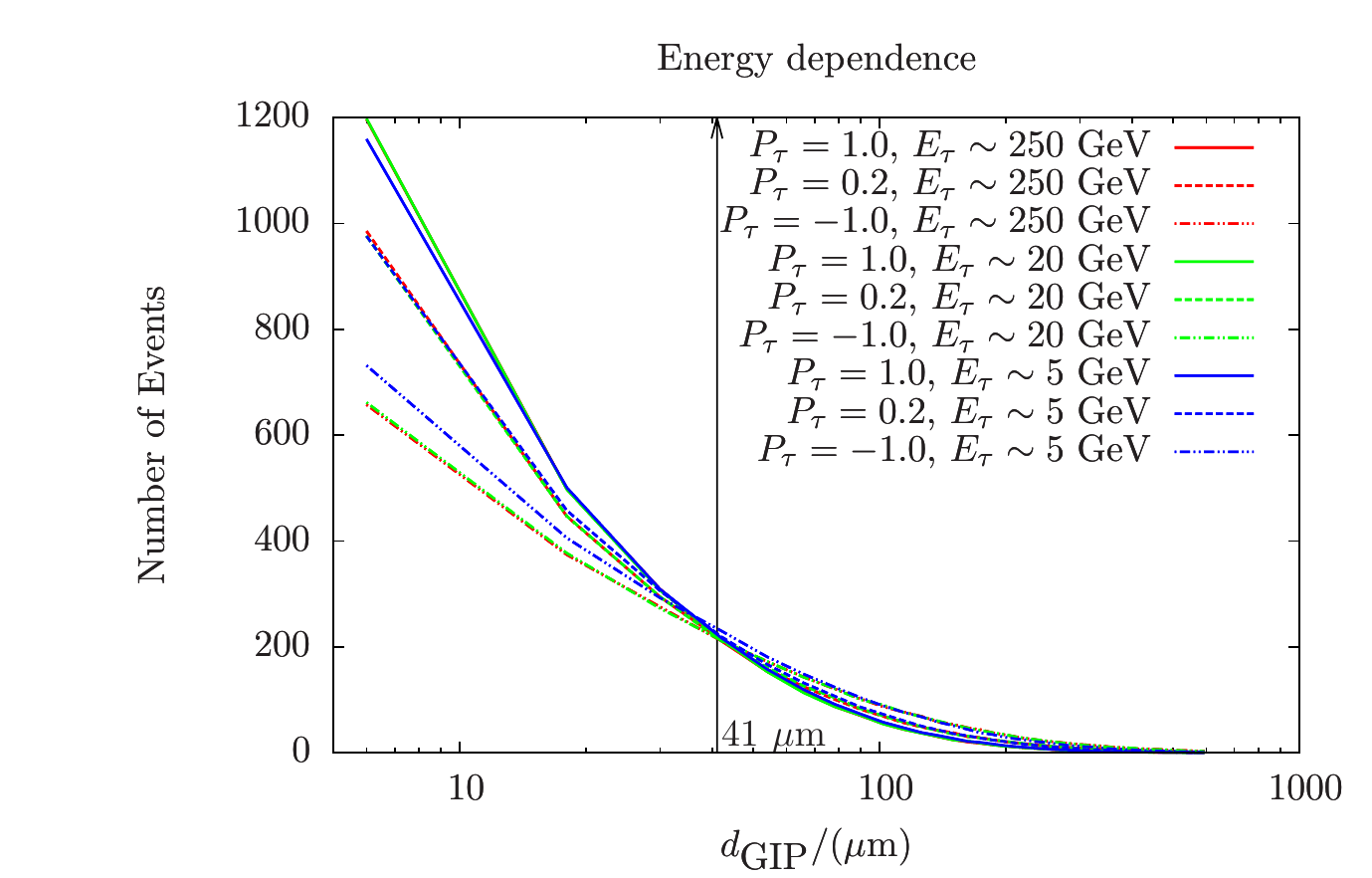} 
    \includegraphics[width=0.48\textwidth]{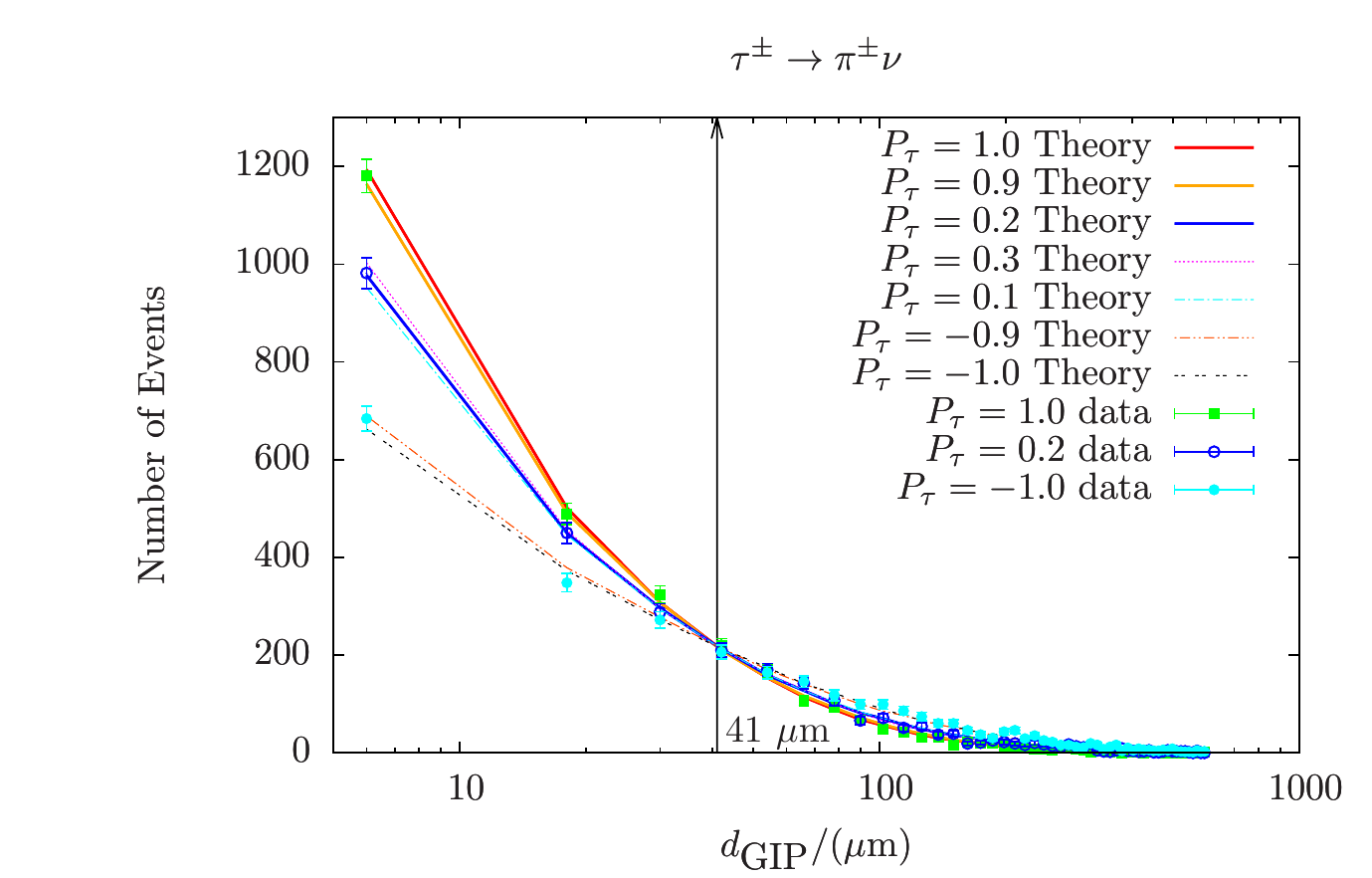} 
    \caption{\label{gip} The polarization dependence of $d_{\text{GIP}}$ for given different energies (left panel) and for a concrete process with $m_{\tilde{\tau}}=300$ GeV and $m_{\tilde{\chi}^0}= 50$ GeV (right panel). } 
\end{figure}

Finally, we propose to use the information from the impact parameter of the pion track in the $\tau \to \pi \nu$ channel to study the tau polarization. 
This variable is more easily accessible experimentally than $z=E_{\pi^\pm} / E_\rho$.  
The decay length of the tau is generated with the probability distribution $P = e^{-L/( \gamma c \tau)}$. Hence, we can calculate the geometrically signed impact parameter by using Eq.~(\ref{eqgip}). 
The distribution of the geometrically signed impact parameter $g_{\text{GIP}}$ is shown in Fig.~\ref{gip}, where the curves for all polarizations of taus appear to be intersecting at around $d_{\text{GIP}}\sim 41 \mu \text{m}$.
So we can define 
\begin{align}
A_{\text{GIP}}=\frac{N_{|d_{\text{GIP}}|<41} - N_{|d_{\text{GIP}}|>41}}{ N_{|d_{\text{GIP}}|<41} + N_{|d_{\text{GIP}}|>41}}
\end{align}
as a measure for studying the tau polarization for this variable, where $N_{|d_\text{GIP}|<41}$ is the number of events with
$|d_{\text{GIP}}|<41\mu\text{m}$. Note that the modulus is necessary since the definition of $d_\text{GIP}$ allows for it to be negative.
A remarkable feature of $d_{\text{GIP}}$ as shown in the left panel of Fig.~\ref{gip} is its insensitivity to the energy of the tau lepton. A simple
explanation of this is that for an energetic tau, the decay length extension due to the additional boost along the tau momentum direction is canceled by the narrowed angle difference between pion and tau.
As a result, for tau energy larger than $\mathcal{O}(10)$ GeV, the distribution of $d_{\text{GIP}}$ is strongly dependent on the tau polarization and very weakly dependent on the tau energy. 
For tau energy lower than $\sim 5$ GeV,  the energy dependence of $d_{\text{GIP}}$ increase. In this region, the tracks from tau decay tend to have energy smaller than 1 GeV which can not be reconstructed effectively at ILC and tau identification may also suffer from heavy contamination. However, as we can see from Fig.~\ref{etau}, there are few events with tau energy in this region.

\begin{figure}[htb]
  \centering
    \includegraphics[width=0.48\textwidth]{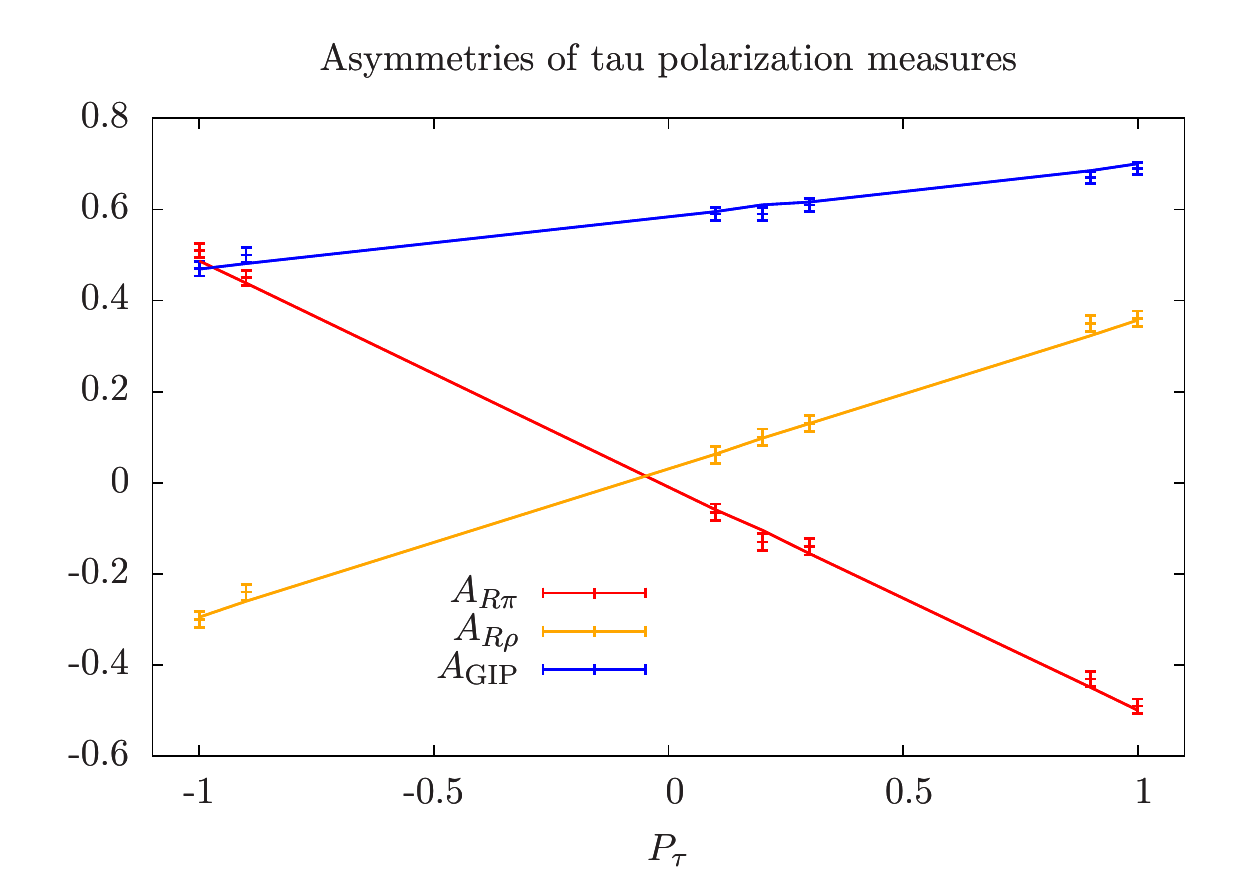} 
    \caption{\label{bench} The three asymmetries as function of the tau polarization $P_\tau$. The lines and points correspond to ``theoretical prediction" and ``experimental measurement" as defined in the text. In case of $A_{R\pi}$, the tau energy has been cheated from MC truth.} 
\end{figure}

Having defined measures for all those variables, we are able to study their relative sensitivities to tau polarization. 
Firstly, we will calculate the corresponding measure for each polarization with a  large number of simulated events, where the statistical uncertainty is very small. We refer this value as the ``theoretical prediction". 
We also consider measures for a much smaller Monte Carlo data set (containing 3000 tau decays) which is  referred  as the ``experimental measurement".  
Then, the corresponding statistical uncertainty for the smaller data set can be calculated by using Eq.~(\ref{staerr}).  
The results are presented in Fig.~\ref{bench}.

From Fig.~\ref{bench} we find that the experimental measurements agree with the theoretical predictions within uncertainties in all cases. 
For a given $P_\tau$, the sensitivity of each measure is proportional to the ratio between the slope of the asymmetry from the theoretical prediction and the uncertainty of experimental measurement at $P_\tau$. 
Since the differences in the slopes of asymmetries are larger than the differences in uncertainties, the sensitivities can be simply estimated as the magnitude of slopes. 
Then the asymmetry $A_{R\pi}$ of the $x$ distribution is giving the most sensitive probe as expected, but recall that it is not experimentally reconstructable. The asymmetry $A_{\rm GIP}$ of the impact parameter is less efficient than asymmetry $A_{R\rho}$. 
On the other hand, in practice $\rho$ meson tagging is generally used to identify the $\tau \to \rho \nu$ signal and to suppress backgrounds. However, this will weaken the sensitivity of $A_{R\rho}$.
A tau decaying into a single charged $\pi$ has a higher tagging efficiency than a tau decaying into a $\rho$ meson. All those considerations suggest that the impact parameter of $\tau \to \pi \nu$ decay as a very promising measure for the study of tau polarization.

\section{Simulation and results}
\label{sec:num}
In the previous section, we have reconstructed the new particle mass and spin by using extra information from the secondary decay of the tau lepton. We also propose a new method using the impact parameter to measure the tau polarization which can provide information about the new physics coupling. 

In a realistic experiment, we have to recognize the features of our signal processes so that the corresponding backgrounds can be specified. A relatively pure signal sample can be collected after the backgrounds are subtracted. Moreover, a detector can only have finite resolution on energy and direction measurement, which will lead to further complications. As has been seen in Sec~\ref{sec:npr}, we will  encounter quadratic equations when we try to solve the system. 
The effects of finite resolution may lead to complex solutions for those quadratic equations, which means that the reconstruction
has failed. 
In what follows in this section, the background subtraction, NLO effects and detector smearing effects will be discussed. 

\subsection{Backgrounds and parton shower effects}
Our signal processes are featured by two tau leptons and relatively large missing energy ($\gtrsim 50$ GeV) in the final state. The channels of interest are $\tau_h \tau_l + \slashed{E}_T$ and $\tau_h \tau_h + \slashed{E}_T$. Our study requires at least one hadronically decaying tau. The pure leptonic decay mode will not be of interest to us.  

There are many SM processes that could potentially contribute backgrounds in our analysis. 
However, some of them can already be understood to be small with appropriate treatment.  SM processes with quark final states can be effectively suppressed by requiring low track multiplicity on jets~\cite{Schade:2009zz}. It is known
that the $\gamma \gamma$ background is negligible when the tau lepton becomes energetic~\cite{Bechtle:2009em}. In addition,
a moderate cut on missing transverse momentum $\slashed{P}_T > 15$ GeV can greatly reduce the $eell$ backgrounds~\cite{Nojiri:1996fp}. 

\begin{figure}[htb]
  \centering
    \includegraphics[width=0.48\textwidth]{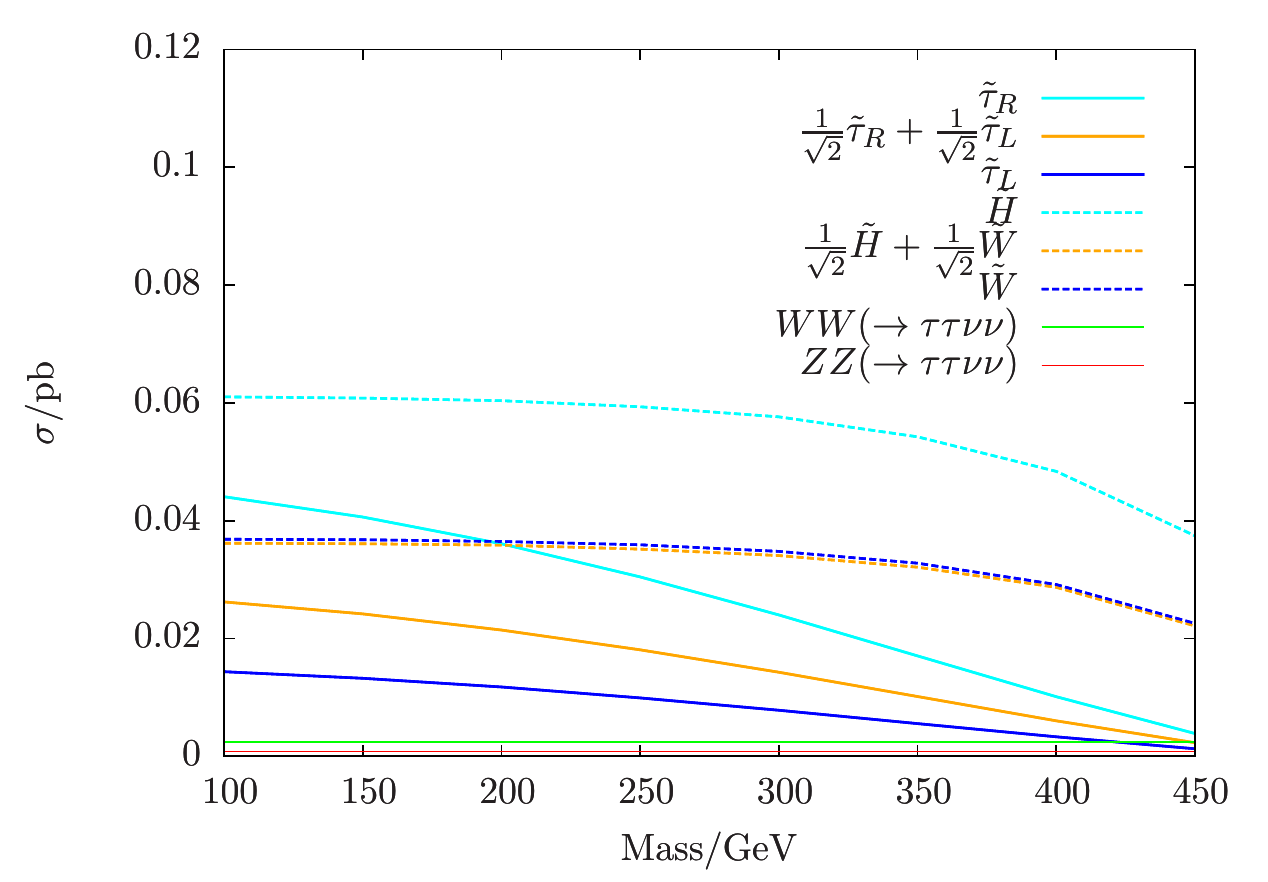} 
      \includegraphics[width=0.48\textwidth]{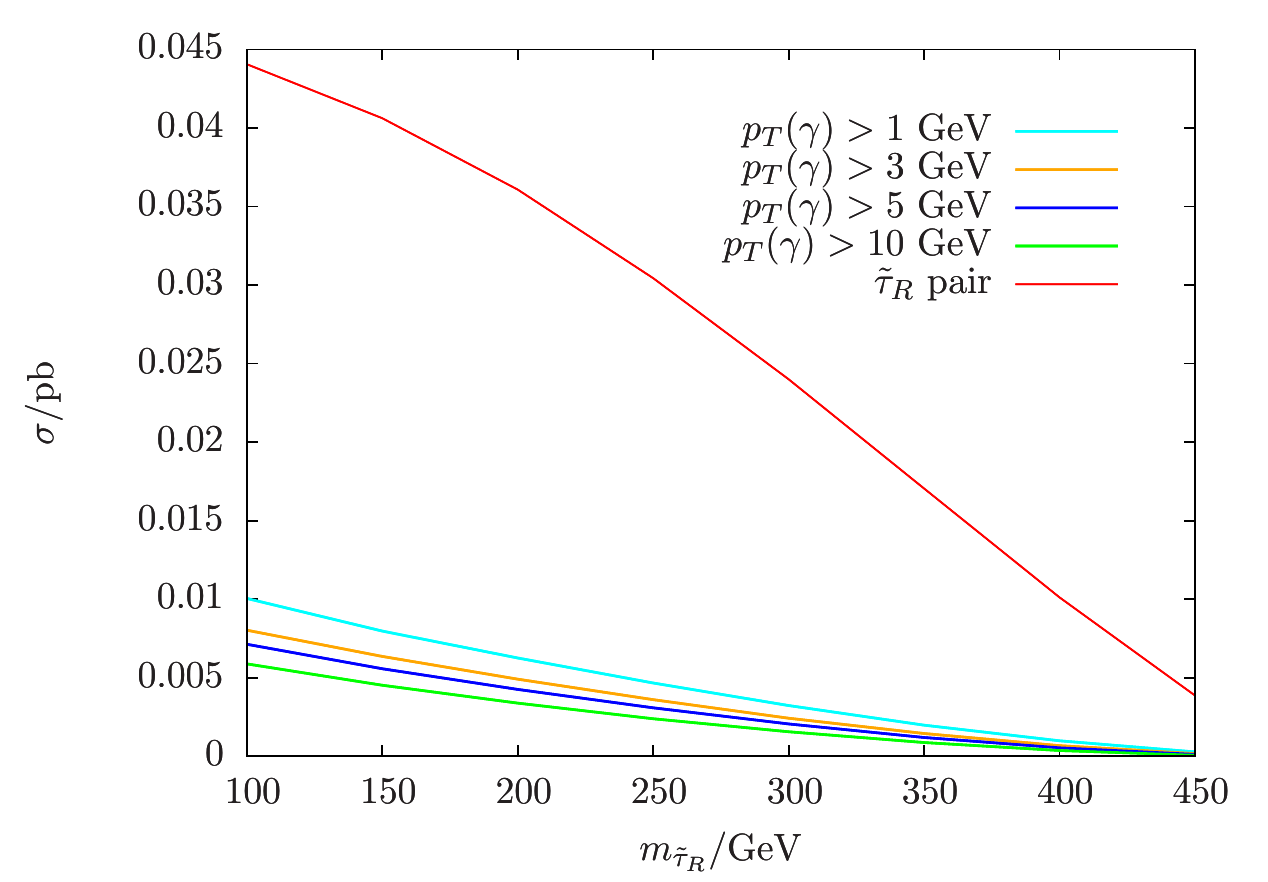} 
    \caption{\label{xsec} 
Cross sections at a 1 TeV $e^+e^-$ collider for an electron beam with 0.8 polarization and a positron beam with
-0.2 polarization.
The left panel shows the production cross section for signals and backgrounds. Three different stau (chargino) mixings are considered. 
The right panel gives the $\tilde{\tau}_R$ pair production cross section with varying cuts on the transverse momentum of the initial
state radiation (ISR) photon.} 
\end{figure}

As a result, the backgrounds of our signal at an $e^+ e^-$ collider are dominated by $WW \to \tau \nu \tau \nu$ and $ZZ \to \tau \tau \nu \nu$. The $Z$-boson mediated $s$-channel and electron neutrino mediated $t$-channel contribution to $WW$ production and the electron mediated $t$-channel contribution to $ZZ$ production can be highly suppressed by using a right-handed polarised electron beam in the collision.  
The corresponding production cross section for beam polarizations $P_{e^-} = 0.8$ and $P_{e^+} =-0.2$  is shown in the left panel of Fig.~\ref{xsec}. Some representative cross sections for processes of interest are also shown.
Those LO cross sections are calculated using MadGraph5~\cite{Alwall:2011uj} with the gauge bosons decaying into taus and neutrinos. The tau lepton in the final state is required to have $p_T>10$ GeV and $|\cos \theta| < 0.98$.  As can be seen from the figure, the $WW$ backgrounds are smaller than the signal processes of interest by more than one order of magnitude in most cases, while the $ZZ$ background is always negligible. 

As for NLO effects, those real corrections (radiation of an extra photon) can effect our reconstruction methods proposed in Sec.~\ref{sec:npr}, while virtual corrections will only lead to an overall normalization. 
For the $P_\tau$ measurement the effects of initial state radiation (ISR) lead to a reduced $\sqrt{s}$  and can be described by a correction factor~\cite{Schade:2009zz}.  
In the following, we will show the smallness of such effects in much simpler way. 
At 1 TeV ILC, the typical $\tilde{\tau}$ energy is $\sim500$ GeV. The radiation of a photon with energy $\sim \mathcal{O}(1)$ GeV can only affect the stau energy by $\lesssim 1$\%. 
Moreover, the production cross sections of a purely right-handed stau ($\tilde{\tau}_R$) pair with varying cuts on the transverse momentum of an ISR photon are presented in the right panel of Fig.~\ref{xsec}. From the figure, we observe that there are no more than 10\% of events with an ISR photon energy larger than $\sim \mathcal{O}(1)$ GeV. This leads us to conclude that photon radiation effects are negligible comparing to detector resolution that will be discussed later. 

From the discussion above we observe that the backgrounds and the NLO corrections are around one order of magnetite smaller than the LO signal process. So, we can safely  discuss our method at LO order without considering any of those effects. Moreover, other features of our signal  may help to suppress the backgrounds  even further, while keeping our signal  intact, e.g. $\slashed{E}_T > 50$ GeV.  However, the precise value of such a kinematical cut depends on the model being considered
and is beyond the scope of our current work. 

\subsection{Detector effects}

With the arguments that known backgrounds and NLO effects can be ignored we are then
ready to discuss the effects of finite detector resolution~\footnote{A comprehensive discussion of detector effects for stau search at ILC can be found in Ref.~\cite{Schade:2009zz}}. 
Let us first note that the tau momentum direction is determined by the direction from the IP to the reconstructed secondary vertex. Its precision is mainly limited by the beam bunch length along the $z$ axis~\footnote{When there is a small crossing angle($14$ mrad) between two beams at ILC, a narrow beam bunch might allow us to reconstruct the IP with precision of $\mathcal{O}(10)~\mu$m along $z$ direction.  }. 
The position of the IP can be calculated using a technique similar to that proposed in Ref.~\cite{Elagin:2010aw}. 

\begin{figure}[htb]
  \centering
    \includegraphics[width=0.48\textwidth]{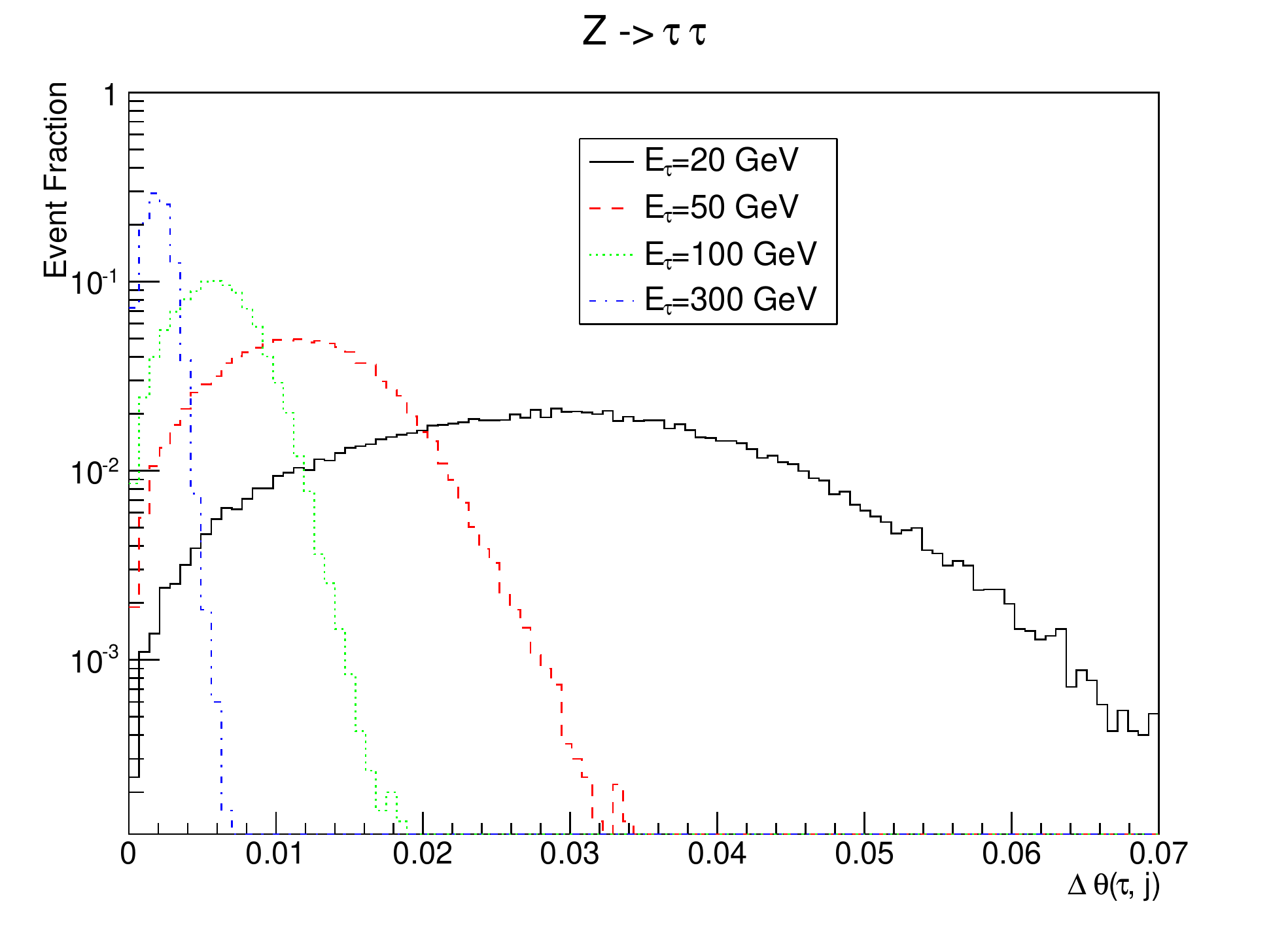} 
    \caption{\label{da} Distribution of the separation angle between $\tau$ and its visible final states for different tau energies.} 
\end{figure}

Firstly, we calculate the probability distribution of the separation angle between the $\tau$ and total visible final states for given $\tau$ energies, by using the process $e^+ e^- \to Z \to \tau^+ \tau^+$ with fixed center of mass energy. Note that  only the 3-prong tau decay channel is of concern for mass and spin reconstruction in this work. As we can infer from Fig.~\ref{er}, the $E_j/E_\tau$ distribution and thus the separation angle distribution for 3-prong tau decay  is not sensitive to the tau polarization. 
So, the probability distribution $P_{\tau}(E_\tau, \Delta \theta(\tau, j))$ deduced from above process can be used for all other processes with arbitrary tau polarization. 
Moreover, the $e^+ e^- \to Z \to \tau^+ \tau^+$ process has another two advantages:
the energy of the $\tau$ from this process is simply equal to the beam energy and it has a large production rate at the ILC which enables precise construction of $P_{\tau}(E_\tau, \Delta \theta (\tau, j))$.
The separation angle distributions for different $\tau$ energies are given in Fig.~\ref{da}.

Secondly, in our case, because of the narrow beam size of the ILC, we only have one degree of freedom, which is the IP position
along $z$ axis, i.e., we denote the reconstructed IP as being at $(0,0,z)$ with the real IP defined to be at $(0,0,0)$. For a given $z$, with the information available from the visible $\tau$ decay products, we can solve for both the $E_\tau$ and the $\Delta \theta(\tau, j)$.  Note that there is a two-fold ambiguity of $E_\tau$ and the one that provides larger $P_{\tau}(E_\tau, \Delta \theta(\tau, j))$ is chosen. Two different likelihood functions are defined, $P_1(z) = P_{\tau}(E_\tau, \Delta \theta(z))$ for a single 3-prong tau in mass reconstruction and $P_2(z) = P_{\tau_1}(E_{\tau_1}, \Delta \theta(z)) \times P_{\tau_2}(E_{\tau_2}, \Delta \theta(z)) $ for two 3-prong taus in spin reconstruction. The $z_{\text{max}}$ which maximize the $P_1(z)$ or $P_2(z)$ is used as the position of the reconstructed IP for each event.

\begin{figure}[htb]
  \centering
    \includegraphics[width=0.48\textwidth]{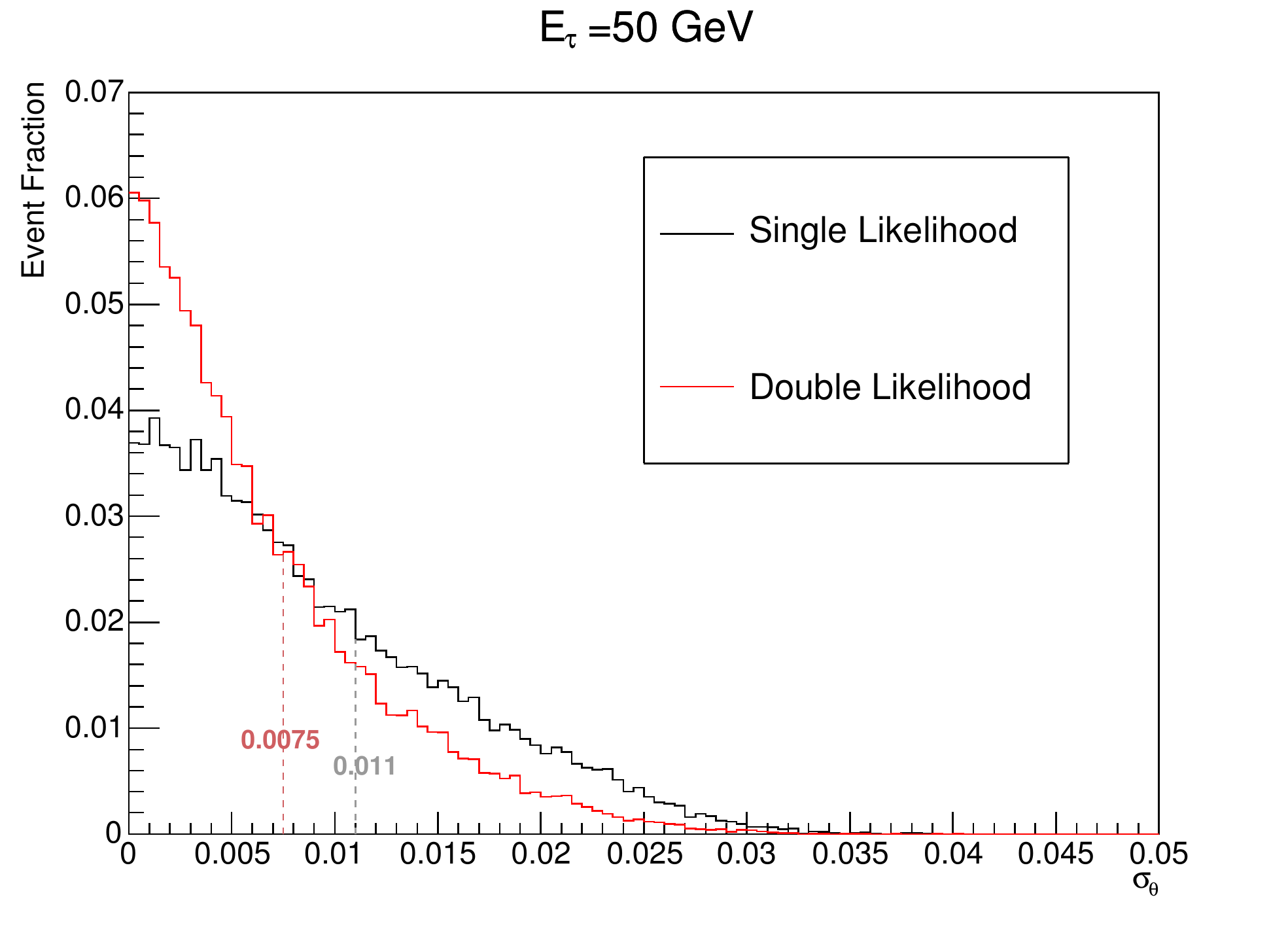} 
    \includegraphics[width=0.48\textwidth]{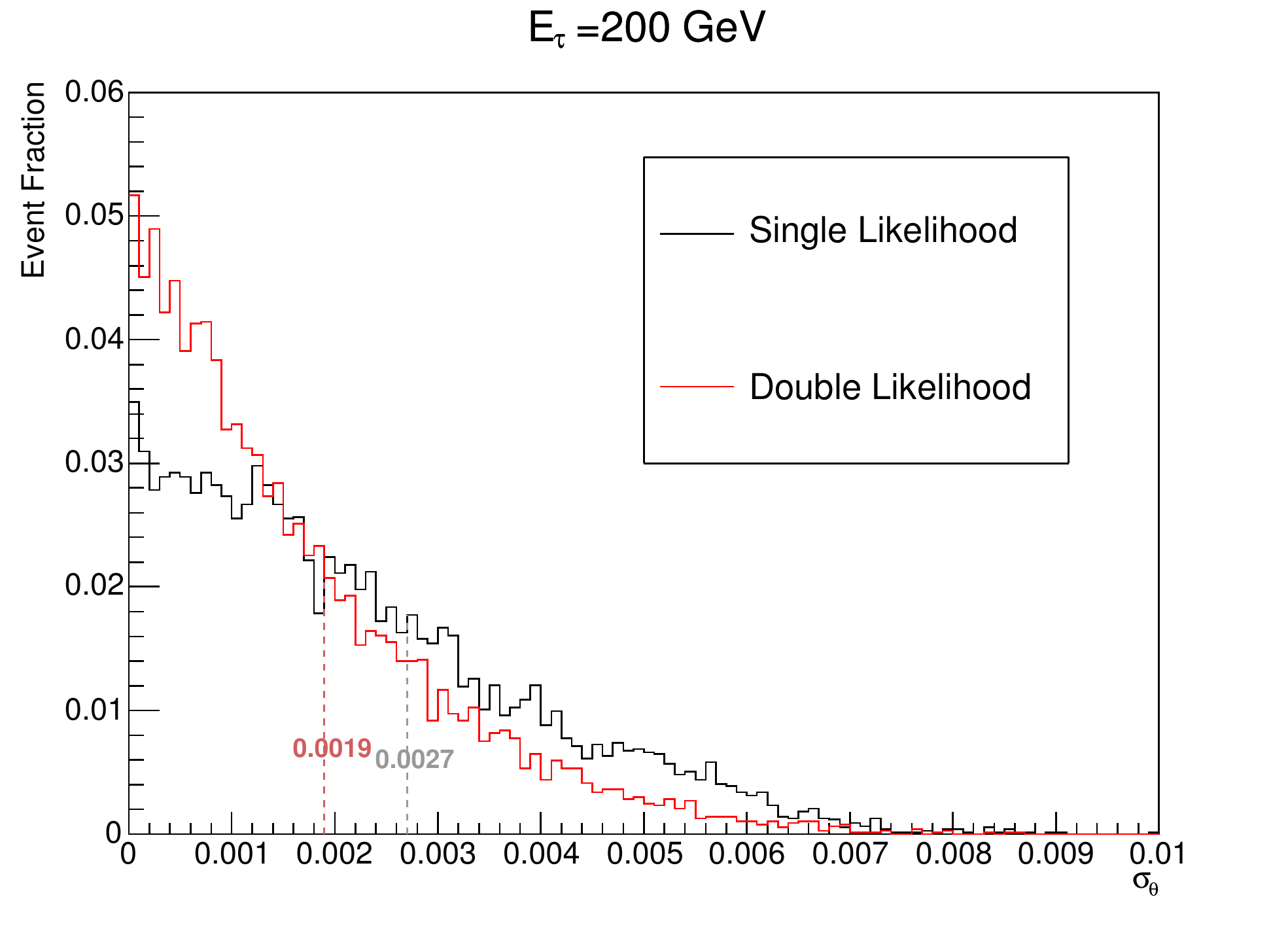} 
    \caption{\label{taua} Angular resolution for three prong decaying tau with $E_\tau=50$ GeV (left panel) and $E_\tau=200$ GeV (right panel). The black curve is for the single likelihood $P_1(z)$ and the red curve is for double likelihood $P_2(z)$.  } 
\end{figure}

In order to estimate the uncertainties of the above reconstruction,  we select our $\tilde{\tau}$ pair events with the $\tau$ lepton energy chosen in two ranges as an illustration, i.e.,  [49.5,50.5] GeV and [199.5,200.5] GeV. 
The results are given in Fig.~\ref{taua}, where the $\sigma_\theta$ is the angular separation between $\vec{\tau}_o = (L \sin \theta_\tau \cos \phi_\tau, L \sin \theta_\tau \sin \phi_\tau, L \cos \theta)$ and $\vec{\tau}_r = (L \sin \theta_\tau \cos \phi_\tau, L \sin \theta_\tau \sin \phi_\tau, L \cos \theta-z_{\max})$. 
We can conclude from the figure, for the single likelihood $P_1(z)$, that the reconstructed tau direction is centered on the true tau direction, with 1-$\sigma$ deviation approximated by $\frac{0.5}{E_{\tau}}$. This result is also consistent with Fig.~\ref{da}, where the angular separation distribution of tau decay indeed shows the  width around $\frac{0.5}{E_{\tau}}$. As for the double likelihood $P_2(z)$, the combinational  effect can improve the angular resolution to $\frac{0.37}{E_\tau}$.

%In the later case, both tau are required to have 3-prong hadronic decay, and only one degree of freedom along the $z$ direction. 

%A more delicate analysis can be used to achieve a better resolution of the position of IP~\cite{Elagin:2010aw,Wasserbaech:1998ur}. That is the reason we used a better precision for the tau direction in this case.  

All other detector effects considered in this work are listed as follows~\cite{Behnke:2013lya}:
\begin{itemize}
\item The tracks used in following analysis are required to have $E_{\text{track}} > 1$ GeV and lie in $|\cos \theta|<0.98$;
\item Energy smearing for hadronic tau jets is taken as $\frac{\Delta E}{E} = 3 \%$;
\item Impact parameter resolution in $r\phi$-plane is $5 \mu \text{m} \oplus \frac{10 \mu \text{m}}{p[\text{GeV}] \sin^{3/2} \theta}$; and
\item The tau identification efficiency is assumed to be 0.7.  
\end{itemize}

From Fig.~\ref{xsec}, we determine that the production cross section for the signal process should be at least $\sim 10$ fb. We will take the worst case and work at a 1 TeV $e^+ e^-$ collider with 1000 fb$^{-1}$ integrated luminosity to illustrate the new physics reconstruction with detector resolution effects included.

\subsubsection{Mass reconstruction}
In order to reconstruct the masses  of the new physics particles, we will need at least one tau that goes through 3-prong decay. As a result, the corresponding number of 3-prong tau decays after tau tagging  and branching ratio suppression is
\begin{align}
N_m \sim 10 fb \times 1000 fb^{-1} \times 0.7^2  \times (13\% \times 2  ) \sim 1200~.~
\end{align}

\begin{figure}[htb]
  \centering
    \includegraphics[width=0.48\textwidth]{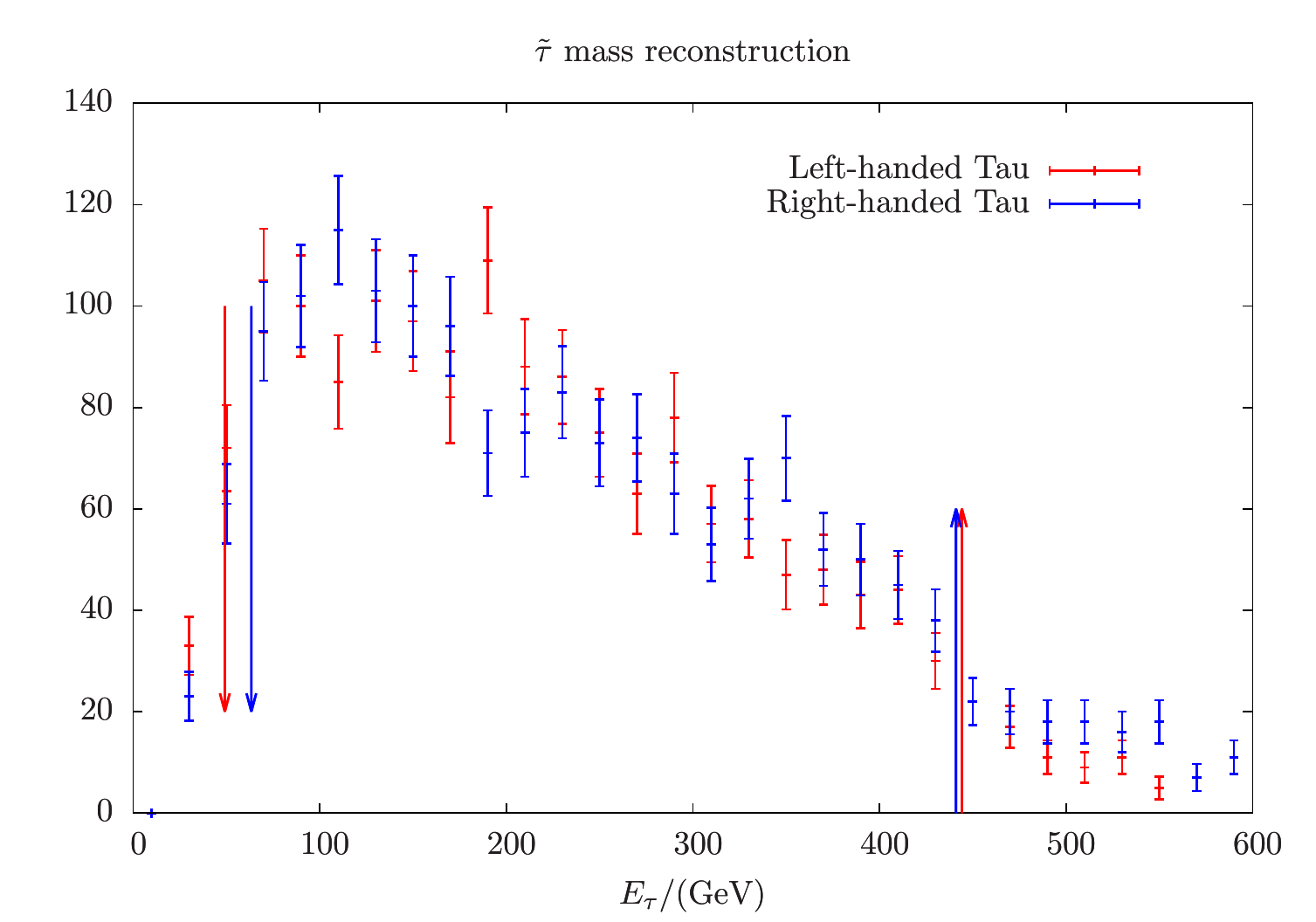} 
    \caption{\label{exmass} The full tau energy distribution for $m_{\tilde{\tau}} =300$ GeV and $m_{\tilde{\chi}^0}=50$ GeV with finite detector resolution. The arrowed lines show the reconstructed endpoints ($E_{\max/\min}$) for $E_{\rm true}$ distributions. }
\end{figure}

The tau energy for mass reconstruction is given by Eq.~(\ref{etaumain}). A finite resolution will lead to negative square root in some cases, which means the reconstruction has failed and so the event should be dropped. With the resolution parameters given above we find around 1/3 of the total events fail the reconstruction.  So we are left with $\sim 900 $ three-prong tau decays for mass reconstruction. The superposed $E_{\text{true/false}}$ distribution for those events are displayed in Fig.~\ref{exmass}. 

Due to the smaller number of events and the smearing effects of detector, the rising and falling edges are spread over more bins than the ones discussed in Sec.~\ref{pmass}. Therefore information from more bins needs to be used to locate the edge precisely. The modified algorithm is then described as follows:
\begin{itemize}
\item Five bins are used to locate the falling edge, i.e., find the bin where $(N_{i-2}+N_{i-1}-N_{i+1}-N_{i+2})/\sqrt{N_{i}}$ is maximized;
\item The height of the $E_{\text{true}}$ distribution is estimated by  $h=N_{i-2}-N_{i+2}$;
\item Note that the height of $E_{\text{true}}$  will be slightly underestimated because of the smearing effects.  
So we look for the largest $j$ such that $N_{j+1}-N_{j-1} > h$ (rather than $0.8 h$ used earlier) to locate the rising edge;   
\item Improved estimates of the locations of the edges are given by $E_{\min}= E_{j}+ \frac{S}{2} \frac{2N_{j} - N_{j+1} - N_{j-1}}{N_{j-1}-N_{j+1}} $ and $E_{\max}= E_{i} + \frac{S}{2} \frac{ 2N_{i}+N_{i+1}+N_{i-1} -2N_{i+2} -2N_{i-2} }{N_{i-2}-N_{i+2}}$;
\item The corresponding uncertainties are 
\begin{align}
\delta E_{min} &= \sqrt{(S/\sqrt{12})^2+ \sum_{k=j-1,j,j+1} (\frac{\partial E_{\min}}{\partial N_k} \sqrt{N_k})^2 } \;\;\;\;~\text{and}~ \\
\delta E_{max} &=  \sqrt{(S/\sqrt{20})^2+ \sum_{k=i-2,i-1,i,i+1,i+2} (\frac{\partial E_{\max}}{\partial N_k} \sqrt{N_k})^2 .}
\end{align}
\end{itemize}
This gives $E_{\min}=49.2\pm 8.0$ GeV, $E_{\max}=444.2 \pm 6.8$ GeV for the left-handed tau and $E_{\min}=63.4 \pm 6.3$ GeV, $E_{\max}=441.1\pm 8.2$ GeV for the right-handed tau. This corresponds to the reconstructed $m_{\tilde{\tau}} =299.6^{+18}_{-22}$ GeV and $m_{\tilde{\chi}}=34.5^{+16}_{-34.5}$ GeV for left-handed tau
and to $m_{\tilde{\tau}} =320.1^{+20}_{-5}$ GeV and $m_{\tilde{\chi}}=40.0^{+5}_{-40}$ GeV for right-handed tau.

\subsubsection{Spin reconstruction}
To reconstruct the new particle spin, both of the taus in the final state are required to go through 3-prong decays. So the corresponding accessible number of stau pair events is
\begin{align}
 N_s = 10 fb \times 1000 fb^{-1} \times 0.7^2 \times (13\%)^2 \sim 83~.~
\end{align}
We can see from the derivation in Sec.~\ref{sec:npr}, there will be a total of 3 quadratic equations in the reconstruction, which means more severe event loss can be expected from complex solutions. On the other hand as a partial compensation, and as discussed above, a better IP position resolution can be achieved when we have two 3-prong taus. 
Since an event typically has 8-fold solutions, we are free to keep all of the real solutions for an event and drop the ones that are complex. 
Note that we have assumed the input masses uncertainties for $\tilde{\tau}$ and $\tilde{\chi}^0$ to be around 5\% in event reconstruction. 
We find the result that with the inclusion of detector resolution effects that around $\sim 2/3$ events have only complex solutions, 
and we are left with $\sim 30$ stau pair events with one or more real solutions for the study of the new particle's spin. 

Next we study the discriminating power of  $A_s$ after the detector resolution effects are included to establish whether it is still possible to distinguish the spin of the new particle spin using such a limited number of events. 
 Simulating with a very large number of events including detector resolution effects we can determine
that  $A_{\rm fermion}=-0.048$ and $A_{\rm scalar}=0.160$. The corresponding statistical uncertainty can be estimated as $\sqrt{(1-A_{\rm scalar}^2)/N}$. With $30  \sim 1.2^2 \times (1-A_{\rm scalar}^2) /(A_{\rm scalar} - A_{\rm fermion})^2$, we can conclude that the ILC with integrated luminosity of $1000$ fb$^{-1}$ will only be able to distinguish the particle spin by more than  $1\sigma$ for our benchmark process~\footnote{More than $3\sigma$ can be achieved at the final state of ILC with integrated luminosity of 8 ab$^{-1}$~\cite{Barklow:2015tja}}.

\subsubsection{Tau polarization} 
Since we are studying the polarization through $\tau \to \pi \nu$ channel and its decay branching ratio is $\sim 11$ \%, we will have in total 
\begin{align}
N_p \sim 10 {\rm fb} \times 1000 {\rm fb}^{-1} \times 0.7^2 \times (11\% \times 2) \sim 1100
\end{align}
tau decays for our impact parameter analysis. 
One of the advantages of the impact parameter analysis is that it does not suffer from any failed reconstruction problems. Furthermore, the ILC can have a quite good resolution for the impact parameter measurement. 

We first study the detector effects by using a very large number of signal events, since the statistical problem can be studied separately.  The corresponding $A_{\text{GIP}}$ for different polarization of tau after considering the detector effects are given in Tab.~\ref{expol}. 
 
 \begin{table}[htb] 
 \begin{center}
  \begin{tabular}{|c|c c|c c c|c c|} \hline
  $P_\tau$   & 1.0 & 0.9 & 0.2 & 0.3 & 0.1 & -1.0 & -0.9  \\  \hline
  $A_{\text{GIP}}$  & 0.698 & 0.685 & 0.605 & 0.615 & 0.591 & 0.478 & 0.466 \\ \hline
  \end{tabular}
    \caption{\label{expol}$A_{\text{GIP}}$ for different degrees of tau polarization after detector resolution. }
    \end{center}
\end{table}

We can conclude from the table that due to the excellent impact parameter resolution of the ILC, $A_{\text{GIP}}$ works essentially as well as before even with detector resolution effects included.  
A 0.1 variation of $P_\tau$ will typically lead to 0.01 changes in $A_{\text{GIP}}$, by estimating its statistical uncertainty as $\sqrt{(1-A_{\rm GIP}^2)/N}$ we can conclude that 1100 tau decays will be able to provide the measurement of tau polarization within $\sim 0.25$ precision. 

Note that we have assumed full purity of the $\pi$ channel for studying tau polarization. Experimentally the $\pi$ and $\rho$ channel can be distinguished using the energy sum of charged and neutral particles. A dedicated study~\cite{Suehara:2009nj} using neural networks has shown that the purity of the $\pi$ mode can reach 96\%. As for the $d_{\rm GIP}$ distribution of the $\rho$ channel, we find it is almost the same for all different tau polarizations with $A_{\rm GIP} \sim 0.61$, i.e., it is the $A_{\rm GIP}$ 
corresponding to $P_\tau \sim 0$ in $\pi$ the channel. 
So, the small residual admixture of the $\rho$ channel with the $\pi$ channel will act to reduce the estimation of the tau polarization by only a few percent.

\section{Conclusion and discussion}
\label{sec:conc}

In this work we have exploited the relatively long lifetime of the tau lepton
to demonstrate a new way to extract new physics parameters. 

For a tau undergoing a 3-prong decay, its direction can first be determined from the location of its displaced vertex and
subsequently its energy can be determined up to a two-fold ambiguity.
We saw that the distribution of the false tau energy solution of this channel is insensitive to the tau polarization. By extracting the end point energies ($E_{\max/\min}$) from the reconstructed tau energy distribution, we showed that the mass reconstruction
precision of new particles ($\tilde{\tau}$ and $\tilde{\chi}$ in our case) can reach a precision of $\sim 10$ GeV with 3000 three-prong tau decays. 

If we require both tau leptons in the final state to go through 3-prong decay such that we know both tau directions, then the whole system can be reconstructed.  Even though there is an eight-fold ambiguity, we find the distribution of those false solutions have a somewhat flatter shape than the true solution distribution, i.e.,
more concentrated in the central region for the scalar and more concentrated in the forward/backward region for the fermion. By
studying the statistical uncertainty and using the distribution of all of the eight-folds solutions, we find that 
only a relatively  small number of stau pair events ($\sim 120$) are required to establish a $3\sigma$ differentiation between scalar and fermion final states. 

We have also proposed a new method to measure the tau polarization in the $\tau \to \pi \nu$ channel, i.e., by using the impact parameter distribution of the charged pion in the final states. This method has the advantages of being more easily accessed experimentally and of being measurable with high precision at the ILC. We also find the impact parameter distribution is sensitive to the tau polarization while being very insensitive to  the tau energy. This was seen to be particularly true in the parameter
region of interest. We observed that the impact parameter distributions for different tau polarizations appear to
intersect at approximately $d_{\text{GIP}} \sim 41 \mu$m for our signal processes.

With the assumption of a relatively pure right-hand polarized beam, we find that the backgrounds and NLO correction effects
are typically more than one order of magnitude smaller than our leading order process and so these
effects have been are neglected in this first analysis for simplicity. 
Assuming a signal production cross section of 10 fb and taking into account realistic detector resolution effects 
we have found that with integrated luminosity of $1000$ fb$^{-1}$
the mass reconstruction precision can reach $\sim 20$ GeV
for our benchmark point ($m_{\tilde{\tau}}=300$ GeV, $m_{\tilde{\chi}^0}=50$ GeV). 
In attempting spin reconstruction we encounter 3 quadratic equations. Taking into account the effects of finite detector resolution
we find complex solutions in some cases and these must of course be rejected. The rejection rate for false solutions was
found to be much higher than that for true solutions and this leads to a welcome increase in the ratio of
true to false solutions.
This then leads to an increased difference between the superposed distributions for all real solutions for both scalars and 
fermions.
After taking into account all detector effects we conclude that 200 reconstructed stau pair events would be enough to resolve the
spin of a new particle to $3\sigma$ C.L. 
Since the ILC would provide a good resolution for the impact parameter, the discriminating power of $d_{\text{GIP}}$ after
taking into account detecter resolution effects is essentially unchanged. 
The ILC with integrated luminosity of $1000$ fb$^{-1}$ is expected to be able to resolve the tau polarization to within a precision of $\sim 0.25$.

A comparison with some previous studies~\cite{Nojiri:1996fp,Bechtle:2009em,Schade:2009zz} shows that this
earlier work reported better resolution of the stau mass and tau polarization than the results presented above. 
However, this is primarily due to the limited statistics that result from our choice of our relatively heavy benchmark point,
i.e., $m_{\tilde{\tau}}=300$ GeV.
For example, in Ref.~\cite{Bechtle:2009em} there were $\sim 9000$ stau pair events before any selections were used for $\tilde{\tau}_2$ and, in addition, it was assumed there that the 
neutralino mass $m_{\tilde{\chi}^0}$ was already known.
With these assumptions, the end point $E_{\max}$ can be fitted with an uncertainty of $\sim 2$ GeV, which leads
to an uncertainty in $m_{\tilde{\tau}_2}$ of  $\sim10$ GeV. 
Moreover, as pointed out in this reference, the reconstructed $m_{\tilde{\tau}_2}$ is very sensitive to the presumed 
value of $m_{\tilde{\chi}^0}$, e.g., an error of 80 MeV on $m_{\tilde{\chi}^0}$ translates into an additional error of 1.4 GeV on $m_{\tilde{\tau}_2}$. Our quoted lower resolution represents a conservative estimate without assuming
a known value of $m_{\tilde{\chi}^0}$ and also in part results from a smaller statistical sample in the 3-prong tau decay channel of our benchmark point (2600 three-prong tau decays before selection).
As for the tau polarization measurement using the $\pi$ spectrum quoted in Ref.~\cite{Bechtle:2009em},
they used $7.92 \times 10^4$  stau pair events before selection for $\tilde{\tau}_1 \tilde{\tau}_1$ with a corresponding uncertainty on the tau polarization of
$\Delta P_\tau \sim 0.1$, which reduce to 0.06 after considering the correlation between fitted normalization and the polarization.
Assuming the same number of signal events in our study the uncertainty in $A_{\rm GIP}$ is
$\delta A_{\rm GIP} \sim 0.0076$. As can be seen from Table~\ref{expol} the corresponding change in the tau polarization
is $\Delta P_\tau \lesssim 0.1$.
In summary, we see that a similar precision in the tau polarization measurement can be achieved with the same number of assumed events. 

In summary, this work has demonstrated an approach to searches for new physics particles that exploits the
relatively long lifetime of the tau and the resulting displaced secondary vertex of the tau decay. We have shown
that these techniques allow a determination of the mass and spin of the new physics particle. They also 
provide information about the couplings of the new physics particle that can be inferred from measurements of
the tau polarization. This work represents a valuable complementary approach to the determination of these quantities
and the precision obtained is comparable with other approaches.

\section*{Acknowledgements} 
We gratefully acknowledge Yandong Liu and Zhen Liu for helpful discussions.
This research was supported by the Centre of Excellence for Particle Physics at the Terascale (CoEPP), which
is funded  by the Australian Research Council through Grant No. CE110001004 (CoEPP), and by 
the University of Adelaide.

\bibliography{taureco}

\bibliographystyle{utphys}

\end{document}